# Science Challenges in Low-Temperature Plasma Science and Engineering:

# Enabling a Future Based on Electricity Through Non-Equilibrium Plasma Chemistry

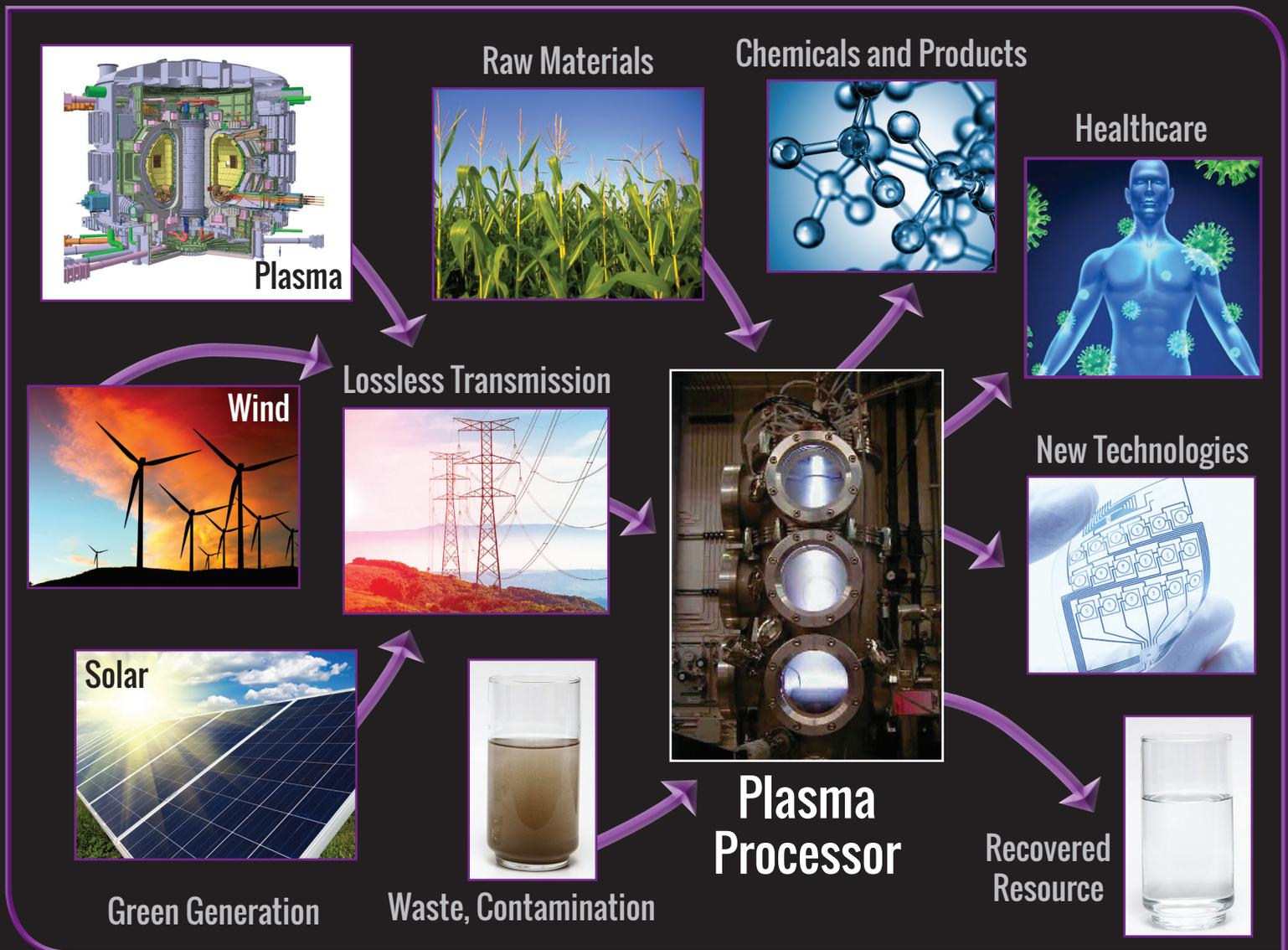

Workshop held at the National Science Foundation | Arlington, VA | 22-23 August 2016

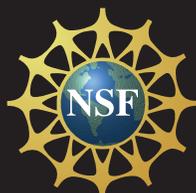

Workshop Report

# Science Challenges in Low Temperature Plasma Science and Engineering:
# Enabling a Future Based on Electricity through Non-Equilibrium Plasma Chemistry

Workshop held at the
National Science Foundation
Arlington, VA
22-23 August 2016

Workshop Co-Chairs:

Selma Mededovic Thagard
Department of Chemical and Biomolecular Engineering
Clarkson University, Potsdam, NY 13699-5705, smededov@clarkson.edu

Mohan Sankaran
Department of Chemical and Biomolecular Engineering
Case Western Reserve University, Cleveland, OH 44106-7217, mohan@case.edu

Mark J. Kushner
Department of Electrical Engineering and Computer Science,
University of Michigan, Ann Arbor, MI 48109-2122, mjkush@umich.ecu

Report Submitted:
Preliminary – February 2017
Final – April 2017

**Acknowledgements**

The attendees and organizers of the workshop, *Science Challenges in Low Temperature Plasma Science and Engineering: Enabling a Future Based on Electricity through Non-Equilibrium Plasma Chemistry*, gratefully acknowledge the support of the National Science Foundation (CBET-1613074) and the Army Research Office (W911NF-16-1-0562). We are also thankful for the advice and support of Dr. Vyacheslav Lukin (PHY), Prof. JoAnn S. Lighty (CBET), Prof. Triantafillos John Mountziaris (CBET) and Prof. Hao Ling (ECCS) of NSF; and Dr. David Stepp and Dr. Michael P. Bakas of ARO. Dr. Maria Burka (NSF) was instrumental in helping form the vision for the workshop and guiding the organizers through the process. The co-chairs are particularly grateful to Prof. David B. Graves (University of California at Berkeley) for his insightful advice on formulating the agenda of the workshop and in producing this report. Dr. Julia Falkovitch-Khain (University of Michigan) was irreplaceable in helping to organize the workshop. Finally, we thank the University of Michigan College of Engineering and the Michigan Institute for Plasma Science and Engineering for their support. Prof. Mark J. Kushner (University of Michigan) compiled and edited this report, aided by Ms. Laura Crane for copyediting and Ms. Rose Anderson for cover design.

**Image credits for report cover**:
- International Thermonuclear Experimental Reactor, www.iter.org.
- Eindhoven University of Technology, Department of Applied Physics. www.tue.nl/en/university/departments/applied-physics/research/plasma-physics-and-radiation-technology/plasma-and-materials-processing-pmp/facilities/technologies/expanding-thermal-plasma-etp.
- CeramTech GmbH, www.ceramtec.com/ceramic-materials/biolox/periprosthetic-joint-infection.
- Engineering Research Consulting, engrrc.com/wp-content/uploads/2016/04/molecules3-1.jpg.
- Fujifilm Europe B.V., green-plasma.eu/index.php/applications-properties/potential-applications.
- BBC Learning, www.bbc.co.uk/programmes/p01z2407.
- The Scientist Magazine, www.the-scientist.com/?articles.view/articleNo/32594/title/A-Win-for-GM-Crops.
- Bigstockphoto.com.

# Table of Contents

## I. Executive Summary

The National Academies describe global challenges as society-level priorities requiring international collaboration to innovate solutions [1]. Perhaps the greatest challenges are centered on energy and environment – collectively called sustainability. A rapidly converging vision of the future portrays societies sustained by green electricity generated by renewable resources. To enact this vision of a *future based on renewable electricity* (*FBRE*), sustainable power must be harnessed at large enough scales to produce the essential chemical reactivity that fuels modern society. This goal aligns with the National Science Foundation (NSF) vision in which these grand challenges are described, in part, as "protecting human health; understanding the food, energy, water nexus" [2]. The NSF rightly cites the importance of convergent research in achieving these goals. This report summarizes the role of low temperature plasmas (LTPs) and the LTP science challenges that must be met to achieve the goal of the *FBRE*.

LTPs are partially ionized gases composed of neutral particles, radicals, excited states, ions, and electrons, the latter of which have temperatures of a few to 10 eV (1 eV = 11,600 K). Low temperature means that while the electrons are at high temperatures, the atoms, molecules and ions of the plasma are typically close to room temperature. In LTPs, power transfer from electrons to atoms and molecules efficiently produces activated species (e.g., radicals, excited states, photons) and chemical reactivity. With such properties, LTPs are essential to technologies ranging from microelectronics to surgical tools.

The science and technology of LTPs harbor dynamic and versatile methods of converting the potential energy of electricity to chemical reactivity, thereby enabling the *FBRE*. Research on LTPs connects fields as diverse as engineering, plasma physics, biology and medicine, and so *LTPs embody the definition of convergent research.* This research will give rise to sustainable products, carbon neutral chemicals, medical advances, recovered resources, advanced materials, improved food and water security, and environmental stewardship.

*LTP science, if properly stewarded, has the potential to develop technologies capable of converting electricity into chemical reactivity and new materials at the scale, efficiency, and selectivity required to meet the needs of a rapidly changing society in a sustainable way.*

To achieve these goals, significant scientific challenges must be addressed and a programmatic home for LTPs established. The workshop, *Science Challenges in Low Temperature Plasma Science and Engineering: Enabling a Future Based on Electricity through Non-Equilibrium Plasma Chemistry*, was held at the NSF in August 2016. The attendees developed a roadmap reflecting the highest impact, highest return scientific challenges in LTPs in the context of controlling chemical reactivity for a sustainable future. This report summarizes their findings. Emphasis was on four focus application areas:

- Multiphase Plasma Systems
- Biotechnology and the Food Cycle
- Energy and the Environment,
- Synthesis and Modification of Materials

The field of LTPs is perhaps unique in being able to impact the broad intellectual diversity represented by these areas, and this diversity represents the convergent nature of the field. This same intellectual diversity makes it difficult to condense the science challenges of the field into a few sentences. However, there are unifying themes which transcend the field in the context of the *FBRE*:

- *Plasma Produced Selectivity in Reaction Mechanisms in the Volume and on Surfaces:* Selectivity is the basis of the chemical and materials industries. The scientific challenge that unites the field is devising methodologies whereby plasma produced chemical selectivity can be improved based on knowing the molecular properties of the feedstocks and the desired products. Meeting this goal requires improving our fundamental understanding of plasma particle distributions, plasma-surface interactions and plasma-wave interactions.

- *Interfacial Plasma Phenomena – Surfaces, Interfaces and Nanostructures:* LTP applications often involve interaction of plasmas at *multiphase boundaries* (e.g., gas-liquid) and the transport of activated species to and through the interface. Investigating and mastering the fundamental processes of LTPs intersecting with multiphase boundaries, from liquids to organic tissues, will enable unprecedented advances in the development of new technologies.

- *Multiscale, Non-Equilibrium Chemical Physics – Emergent Plasma Phenomena:* LTPs are intrinsically non-equilibrium which enables production of unique chemical reactivity. *Collective phenomena emerge* from interactions between the individual non-equilibrium components of LTPs. Controlling and optimizing emergent, collective behavior will be necessary to optimizing the chemical reactivity produced by plasmas.

- *Synergy and Complexity in Plasma:* Multiscale effects in LTPs are dominated by surfaces, interfaces and nanostructures. These combined properties result in system *synergies and complexities* ranging from ion-neutral interactions in plasma etching of semiconductors to the intricate combination of effects in the plasma treatment of biological systems. Mastering these complexities to achieve the goals of a *FBRE* will require new interdisciplinary, convergent approaches that horizontally integrate those disciplines.

Establishing a programmatic home for LTPs would recognize the critical need for convergent research to enable the *FBRE*. Countries in Europe and Asia are vigorously pursuing LTP science through national and international programs. The present model in the US of tackling LTP applications on an individual basis through various Federal programs is not able to address the science challenges set forth in this report. However, a comprehensive LTP program, empowered to pursue the common research challenges, catalyze convergent research and support the fundamental science of LTPs, would enable advances towards the *FBRE* unachievable by other means.

One possible programmatic home for LTPs is the NSF Directorate for Engineering. The authors of this report were asked to address three questions – why are LTPs essential to the NSF vision of a sustainable future, why is engineering the appropriate platform in invest in LTPs, and why is now the time to invest? A synopsis of responses to those questions follows:

- *Why Low Temperature Plasmas?* LTPs represent precisely the convergent science and technology NSF aspires to support to address grand challenge scale problems.
- *Why Engineering (but not only engineering)?* The science resulting from LTP research must be context-driven to rapidly impact sustainability – and that is engineering. However, that focus cannot be only engineering. An LTP program must be empowered to reach out and collaborate with the non-engineering disciplines essential to the *FBRE*.
- *Why Now?* The field of LTPs is uniquely posed to translate the fundamentals of plasma science into society serving applications across multiple disciplines. Given the urgency of the *FBRE*, the need for a programmatic home for LTPs has never been greater.

## II. The Plasma Enabled Sustainable Future

### II.A. The Role of LTPs in Sustainability

Low temperature plasmas (LTPs) represent an integrative field of science and engineering that holds the potential to make unprecedented contributions to sustainable societies. LTPs have the unique ability to activate chemical processes in gases, liquids and solids, and produce outcomes not otherwise possible. The field of LTPs has translated fundamental understanding of plasma generated chemical reactivity into technologies ranging from microelectronics fabrication, human implants, lighting and lasers, to plasma propulsion for interplanetary spacecraft and solar-generated electricity production. This report describes the role of LTPs in addressing a new level of intellectual challenge – a society based on sustainable resources – and the corresponding LTP science challenges and convergent research priorities.

Societies from the developing to the technologically advanced face a daunting set of challenges that must be met to maintain and improve quality of life. The National Science Foundation (NSF) vision of the future identifies these particular grand challenges as "protecting human health; understanding the food, energy, water nexus" [2]. This vision recognizes the need to perform fundamental research as a precursor to new technologies that will protect and improve the food cycle, extend and better utilize finite water resources and develop the means to produce and distribute energy in an environmentally friendly manner. These goals are baseline requirements for a robust quality of life based on state-of-the-art technologies. In many ways, these challenges characterize *sustainability* – developing technologies which use fundamental resources in an environmentally sustainable manner to improve the quality of life. The NSF rightly cites the importance of *convergent research*– "the merging of ideas, approaches and technologies from widely diverse fields of knowledge to stimulate innovation and discovery" [2] – in achieving these goals.

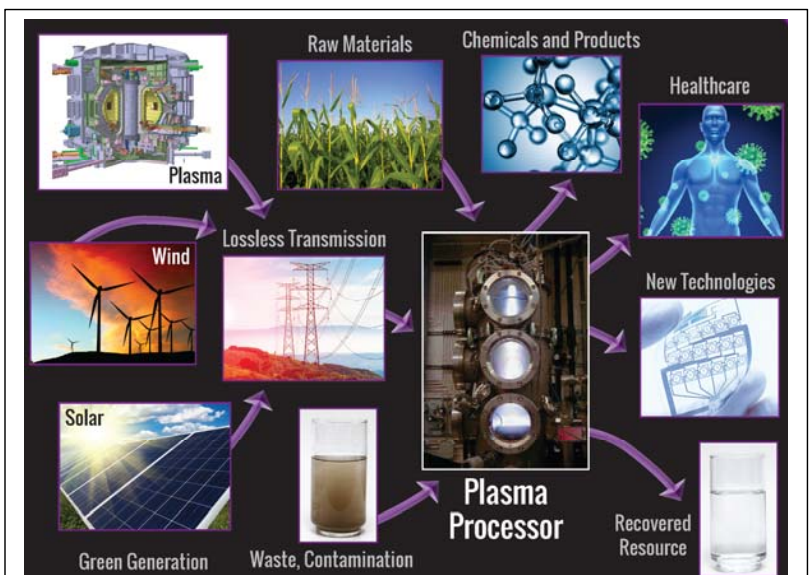

Figure 1 – The *future based on renewable electricity* will be enabled by the plasma processor that converts the potential energy of electricity into the chemical reactivity that fuels society, from recovering resources, new technologies and healthcare, to the products of everyday life. [M. J. Kushner, (2016)]

Sustainability may be the greatest challenge facing society today. From one perspective, this challenge revolves around controlling *chemical reactivity*. Much of the technological foundation of society is based on managing chemical reactivity. Controlled chemical reactivity leads to high-technology materials, fertilizers, plastics, medicines, safe food, clean water and transportation. In short, controlling chemical reactivity drives society – it creates products and manages resources utilized in all of life's activities. Previous breakthroughs in controlling chemical reactivity have transformed society – from combustion for transporta-tion and materials for microelectronics, to water purification and antibacterial agents for disease control. However, we have often been poor stewards when generating and using chemical reactivity. For example, heat engines based on fossil fuel combustion have revolutionized transportation and enabled electrification of societies – but widespread and uncontrolled growth of the technology has led to excessive emissions into the atmosphere.

A rapidly converging vision of the future reveals societies sustained by electricity generated by renewable resources: wind, solar, waves, geothermal, ocean currents, hydroelectric and fusion [3]. In contrast, fossil fuel combustion for power generation, commodity transportation and other non-essential applications is minimized, reserving those precious hydrocarbon fuels for high value transportation and petrochemical-based products that may be impractical to be produced by other means. Most carbon-based products are provided by renewable sources, such as biomass, or by converting low-value, and in some cases, harmful reservoirs of carbon into high-value carbon-containing compounds [4]. Clean, green, sustainable electricity promotes environmental stewardship, protects and expands the food cycle, improves human health and develops new materials needed for advanced societies.

A critical obstacle to enacting this vision is that electricity alone does not produce the essential chemical reactivity that society requires. For a *future based on renewable electricity* (*FBRE*), much of the world's stored energy and power transmitted in the form of electricity must be converted into the energy associated with chemical bonds. There are relatively few ways in which electrical power can be selectively transformed into chemical bonds. One approach is electrochemistry; however, many chemical processes cannot be performed in liquids. A key message of this report is that there is another fundamental approach that has not been significantly explored – plasma chemistry. This report describes how *low temperature plasmas* can be a dynamic and versatile method of converting the potential energy of electricity into chemical reactivity, and so enable the *FBRE*. (See Fig. 1.)

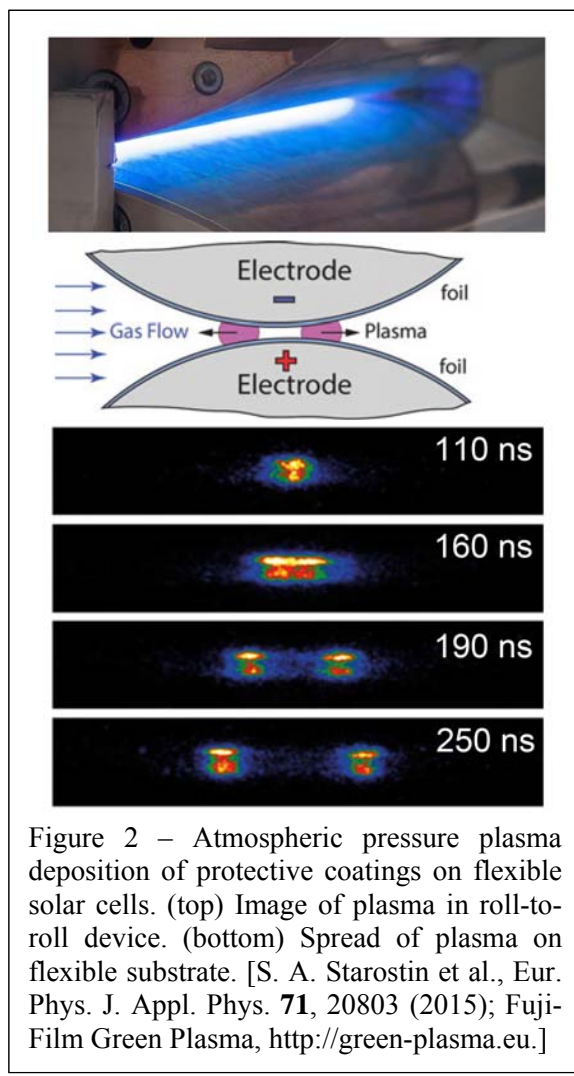

Figure 2 – Atmospheric pressure plasma deposition of protective coatings on flexible solar cells. (top) Image of plasma in roll-to-roll device. (bottom) Spread of plasma on flexible substrate. [S. A. Starostin et al., Eur. Phys. J. Appl. Phys. **71**, 20803 (2015); FujiFilm Green Plasma, http://green-plasma.eu.]

Low temperature plasmas are plasmas associated with electron-volt (eV) science and technologies. LTPs are partially ionized gases composed of neutral particles, radicals, excited states, ions and electrons. They have characteristic electron temperatures of a few to 10 eV (1 eV = 11,000 K) with fractional ionizations that are typically small – less than a few percent to as small as one part per million (ppm). Low temperature in this context indicates that the plasma's neutral atoms and molecules, as well as ions, have temperatures close to room temperature, making them compat-ible with temperature-sensitive materials including living tissue. However, since LTPs have electron temperatures commensurate with the threshold energies of excited states in neutral atoms and molecules, power transfer from electrons to these atoms and molecules efficiently produces activated species – radicals, excited states, photons – thereby initiating and controlling chemical reactions. Acceleration of ions in the boundary layers (sheaths) of LTPs to energies of tens to hundreds of eV activates surface modifying processes such as sputtering, etching and deposition. With such properties, LTPs are essential to technologies ranging from microelectronics [5] to medicine [6]. In multiphase forms, LTPs encompass aspects of basic plasma physics, electrochemistry, photochemistry, catalysis, radiation chemistry, aerosol science and materials science.

*LTP science, if properly stewarded, has the potential to develop technologies capable of converting electricity into chemical reactivity and new materials at the scale, efficiency and selectivity needed to meet the sustainable needs of our rapidly changing society in a sustainable way.*

We envision that LTP-science-based technologies will lead to new devices (flexible solar cells), medical advances (combating antimicrobial resistance), renewable energy production ($CO_2$ conversion), advanced materials (nanomaterials for batteries), chemical fuel synthesis (plasma catalysis conversion of $CH_4$) and solutions to worldwide challenges in food and water (water purification) that will enable the *FBRE* and so contribute to sustainability. (See Fig. 2.) To achieve these goals, the scientific challenges discussed in this report must be addressed and advanced.

## *II.B.     Why LTPs? Why Engineering? Why Now?*

In preparing this report, the authors were asked to address three questions – why are LTPs essential to the NSF vision of a sustainable future, why is engineering an appropriate platform for an investment in LTPs, and why is this the time to invest?

Societies face serious challenges that must be addressed to maintain and improve the quality of life. The need for science-based solutions that rapidly convert to technology has never been greater. And those science-based solutions and technologies will not come from a single discipline working in isolation – the problems are too complex. The NSF perspective for future investments succinctly summarizes this point:

> The grand challenges of today -- protecting human health; understanding the food, energy, water nexus; exploring the universe at all scales -- will not be solved by one discipline alone. They require convergence: the merging of ideas, approaches and technologies from widely diverse fields of knowledge to stimulate innovation and discovery...Convergence blends scientific disciplines in a coordinated, reciprocal way and fosters the robust collaborations needed for successful inquiry. Convergence builds and supports creative partnerships and the creative thinking needed to address complex problems. To build a system that truly supports convergent science, NSF would strategically invest in research projects and programs that are motivated by intellectual opportunities and important societal problems [2].

*Why Low Temperature Plasmas?*

Low temperature plasmas represent precisely the convergent science and technology NSF aspires to support. Historically, LTP science and applications have been interdisciplinary and convergent. The outreach that the LTP community has made to other disciplines has been extraordinary. The LTP community has expanded the fundamental scientific understanding of ion-ized gases and used that knowledge to innovate technologies from lasers, microelectronics, materials processing, and biotechnology to environmental cleanup, high efficiency combustion, medicine, nanotechnology, food sterilization and aeronautical flow control. The LTP community accomplished these extraordinary successes by finding technological needs, embracing and engaging allied disciplines and performing fundamental research to translate science to beneficial technologies. LTP research epitomizes convergence. Indeed, attendees of this workshop represent essentially all engineering disciplines, in addition to physics, materials science, chemistry and biology.

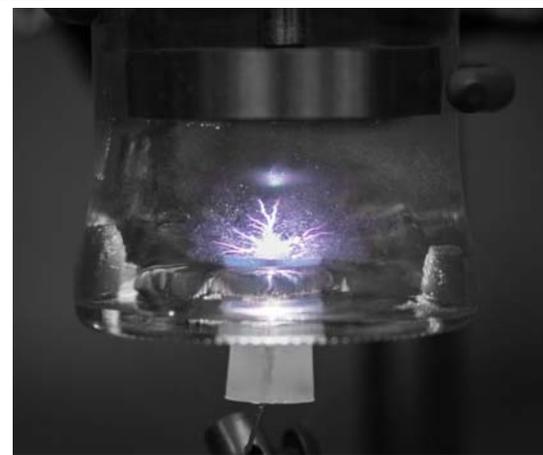

Figure 3 – Plasma water sterilization system for remote, rural use in developing countries that can be produced for low cost from local materials. [S. M. Thagard, Clarkson University, private communication (2017)]

*Why Engineering? (But not only engineering)*

Advances in LTP science will generate critical and unique contributions to sustainably producing chemical reactivity. Many of the fundamental plasma properties recently (and to be) investigated attest to the uniqueness of plasma in producing chemical reactivity in ways that are otherwise inaccessible (e.g., electron-reduction of environmental contaminants at a plasma-liquid interface). However, science advances must be context-driven in order to rapidly impact sustainability – and that is engineering. The dynamic range of potential applications is simply too large (e.g., over nine orders of magnitude in pressure) for a single scientific outcome to universally apply. The science must be performed with some connectivity to the desired outcome – and that is engineering. For example: it is difficult to advance the science of LTPs with the goal of converting $CO_2$ to high valued chemicals without collaborating with the mechanical and chemical engineers implementing the process; it is difficult to advance the science of LTPs with the goal of combating anti-microbial resistance without close connectivity to the biotechnologists and bioengineers whose expertise is essential in evaluating and implementing the methods; it is difficult to advance the science of LTPs with the goal of purifying water absent partnerships with the environmental and civil engineers designing those systems. This is the hallmark of convergent science, and its nexus is engineering.

However, *that nexus cannot be only engineering*. An LTP program in engineering must be empowered to reach out and collaborate with the non-engineering disciplines essential to the *FBRE*. To be successful, convergent research anchored by fundamental scientific advances in LTPs requires close liaisons with programs in basic plasma physics; atomic, molecular and optical (AMO) physics, materials science, biotechnology, agriculture and manufacturing. There is great potential for even further outreach to other disciplines, economic development and humanitarian activities. For example, fundamental research is the basis of a locally-built, low-cost, solar-powered, point-of-use plasma-based water sterilization system designed to serve rural areas in developing countries. (See Fig. 3.)

*Why Now?*

The LTP community has achieved unparalleled success in discovering new science and converting that scientific understanding to societal benefits. This is a remarkable accomplishment considering that the LTP field has never had a home agency. In contrast to other disciplines, en-gineering applications have driven the scientific advances in LTPs. Much of the current understanding of non-equilibrium low pressure plasma science derived from developing applications for microelectronics fabrication. Similarly, current knowledge regarding how high pressure plasmas produce unique chemistries resulted from developing applications for lasers, environmental cleanup and plasma aided combustion. However, the need for an LTP home program has now reached a critical point precisely because this discipline epitomizes convergence. Expecting environmentally focused agencies to support the fundamental plasma science required to achieve their goals is unrealistic. Likewise, expecting a plasma physics program to support the cell culturing required to test whether the plasma has produced the appropriate antimicrobial agent is at best optimistic. These two examples of the convergent nature of LTP research demonstrate the critical and urgent need for a home for LTPs that embraces the intellectual diversity of the field.

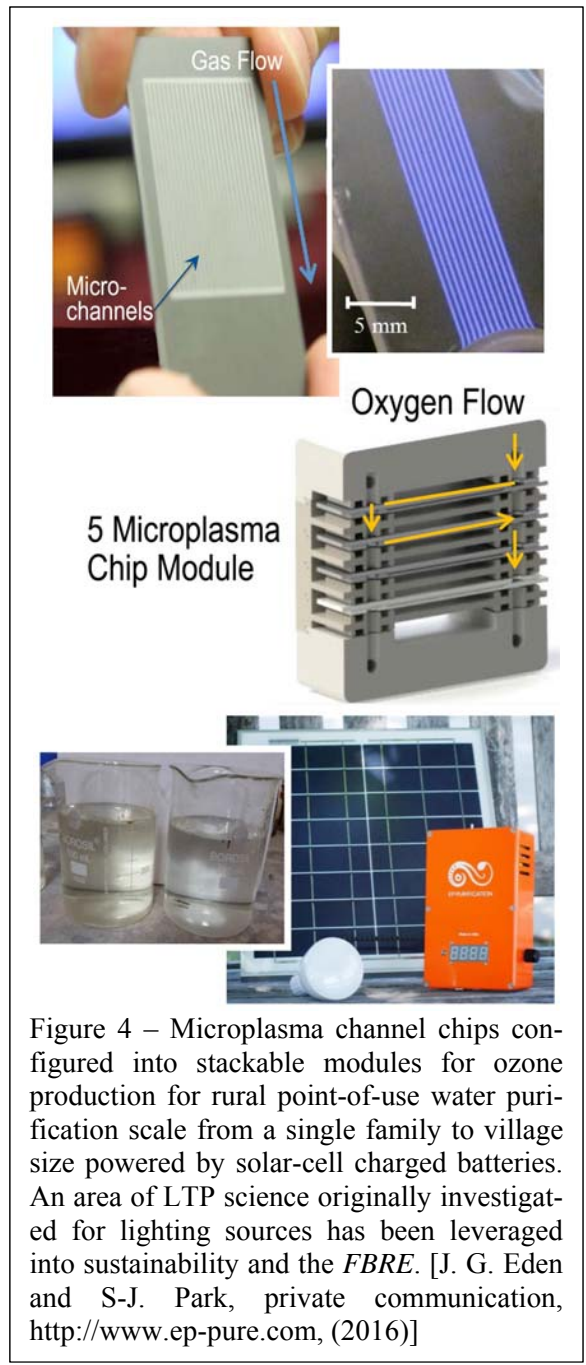

Figure 4 – Microplasma channel chips configured into stackable modules for ozone production for rural point-of-use water purification scale from a single family to village size powered by solar-cell charged batteries. An area of LTP science originally investigated for lighting sources has been leveraged into sustainability and the *FBRE*. [J. G. Eden and S-J. Park, private communication, http://www.ep-pure.com, (2016)]

There are many examples of fundamental LTP research translating to technologies that contribute to sustainability, the *FBRE* and the *food-water-energy nexus*. A specific example addresses the need for rural clean water. Fully 10% of the world's population does not have access to clean water and disproportionately so in rural areas [7]. Given these rural locations, clean water will likely not be delivered by large municipal water systems. The need will be met by village-sized or point-of-use water purification systems. Based on fundamental research on the physics and technology of microplasmas originally intended for lighting sources [8], efficient, modular, arrays of ozone-producing microplasmas have been developed and commercialized for water purification [9]. (See Fig. 4.) These inexpensive compact modules can be scaled by stacking units in a rack for village water treatment, or used individually powered by a battery charged by a solar panel. This technology is now being distributed to developing countries in collaboration with NGOs (non-government agencies) for sustainable point-of-use water purification [10] and being extended to combating higher level contaminated wastewater [11]. The same technology is being investigated for wound treatment in the realm of plasma-biomedicine [12]. This report discusses addressing fundamental science challenges that will build upon such successes.

## II.C  The LTP Workshop

The workshop, *Science Challenges in Low Temperature Plasma Science and Engineering: Enabling a Future Based on Electricity through Non-Equilibrium Plasma Chemistry*, was held at the National Science Foundation in Arlington, VA on August 22-23, 2016. The purpose of the workshop was to develop a roadmap for the future reflecting the highest impact, highest return scientific challenges facing the field of LTPs in the context of controlling reactivity for a sustainable future. Successfully meeting these challenges will advance the knowledge base required for exploiting plasma to achieve the goal of selectively transforming renewable electrical power into products and processes that will fuel those societies. The workshop goals included:

- Summarizing the current state-of-the-art in plasma-activated processes;
- Describing the role of low temperature plasmas in facilitating the vision of a future based on renewable resources;
- Delineating and prioritizing the major scientific issues in LTPs that must be addressed to achieve that vision; and
- Describing a role for NSF in addressing these scientific challenges.

While the workshop broadly addressed the current and anticipated scientific challenges in the entire field of LTPs, particular emphasis was placed on the following four focus application areas which are discussed in-depth in Chapter IV:

- Multiphase Plasma Systems
- Energy and the Environment
- Biotechnology and the Food Cycle)
- Synthesis and Modification of Materials

## II.D.  The Unifying Research Challenges

The focus areas were selected as the outcome of community wide discussions on where plasma produced reactivity will have the greatest influence in addressing sustainability, and the convergent research areas that NSF will focus on in the coming decade – *the food, water energy nexus, and human healthcare*. The scale and dynamics of these focus application areas differ considerably: pressures range over nine orders of magnitude from millitorrs (mTorr) to liquid densities; the bounding materials range from refractory metals to living tissue; and timescales range from picoseconds to days. Defining a single, totally unifying, one-sentence science challenge that captures such intellectual diversity is not possible. However, there are common themes that link not only the four focus areas but the entire LTP discipline. As discussed below, these themes – common high-level science challenges – produce a cohesive and convergent area of research, built upon the unique capabilities of LTPs to produce sustainable reactivity and so enable the *FBRE*.

*Plasma Produced Selectivity in Reaction Mechanisms in the Volume and on Surfaces*

Chemical selectivity is the most fundamental and basic process supporting the chemical, materials and biotechnology industries. From fabrication of solar cells to food processing, thermal equilibrium rarely determines the final state. Rather, a subset of all possible accessible states most often meets the desired outcome. Efficiently producing that desired outcome requires selectivity. Moving towards sustainable societies in which petrochemical feedstocks are replaced by bio-feedstocks, pesticides are replaced by less persistent more targeted alternatives and water reclamation is more energy efficient, will require ever more selective chemical processes with lower energy costs. Low temperature plasmas have the potential to provide this necessary chemical selectivity utilizing renewable electricity. The path towards plasma produced chemical selectivity is a complex one – there will not be a single universal solution that is able to capture the extreme diversity of the field; solutions will likely be system specific. The specific selective solution for plasma based conversion of emission gases such $CO_2$ into high value chemicals will likely be different from plasma activation of aerosols for onsite-production of fertilizers. Having said that, the unifying scientific challenge involves devising a methodology grounded in a fundamental understanding of plasma particle distributions, plasma-surface interactions and plasma-wave interactions, whereby chemical selectivity can be improved based on the molecular properties of the feedstocks and the desired products.

*Interfacial Plasma Phenomena: Surfaces, Interfaces, and Nanostructures*

Many important LTP applications involve the interaction of plasma at multiphase boundaries (e.g., gas-solid or gas-liquid) – that is, interfacial phenomena. Each of the focus application areas include these phenomena, either explicitly (multiphase plasmas with liquids and aerosols; synthesis and modification of materials) or implicitly (energy and environment; biotechnology/food cycle). Critical interactions at phase boundaries very often involve nanoscale interactions, either deliberately, as in the fabrication of nanostructures, or as a natural consequence of the physics, such as the transition zone between gas phase plasmas and liquids. Basic plasma physics typically addresses phenomena within the volume whereas LTP applications, much less studied and understood, are often dominated by plasma interactions with bounding surfaces. It follows then that scientific LTP challenges are intertwined with interfacial phenomena. How do LTPs interact with phase interfaces? How can those interfacial interactions be controlled to produce nanostructures or permeable membranes for water-purification? How can plasma produced reactivity be transferred into liquids for environmental cleanup or to fight antimicrobial resistance. Investigating and mastering the fundamental processes of LTPs intersecting with multi-phase boundaries, from liquids to organic tissue, will enable unprecedented advances in the development of new technologies.

*Multiscale, Non-Equilibrium Chemical Physics: Emergent Plasma Phenomena*

Low temperature plasmas are unique in that they produce intrinsically non-equilibrium conditions. While this non-equilibrium enables LTPs to selectively produce unique chemical reactivity, it also has profound, but still poorly understood consequences. In addition, LTPs often display emergent characteristics – that is, collective effects emerge from interactions between the individual non-equilibrium components. For example, plasma-liquid interactions often display emergent collective dynamics at the interface that can dominate outcomes, from uniformity of processing to interacting with desired phases. This emergent behavior often begins from the selective excitation or production of a specific species that results from capitalizing on the non-equilibrium character of the system. The emergent behavior also may evolve from small to large scales. In applications involving volume processing at atmospheric pressure (e.g., plasma to catalyze chemical reactions for energy and environmental applications), plasma spatial uniformity is often controlled by non-linear collective effects such as formation of interacting plasma filaments. Emergent phenomena can dominate all four focus areas and are a common center of attention for scientific studies of LTP. Controlling and optimizing emergent, collective behavior will be necessary to optimize the chemical reactivity produced by plasmas.

*Synergy and Complexity in LTPs*

Multiscale, non-equilibrium effects in LTPs dominated by surfaces, interfaces and nanostructures result in system *complexity*. An enormous range of phenomena occur simultaneously in LTPs. This synergistic complexity can be exploited in LTP applications such as ion-neutral synergy in plasma etching of semiconductor devices [13]. LTP biological applications involve a bewildering combination of effects. The biochemical effects of reactive plasma produced chemical species (e.g., reactive oxygen and nitrogen) are in turn affected by the presence of plasma generated pulsed electric fields. Mechanical shocks induced by collapsing bubbles created by plasma in liquids, and plasma induced heating or photochemistry can also couple to the electro-chemical effects in altering a biological response. A single plasma process – for example, a plasma jet treating a serum covered wound or a pulsed plasma converting toxic emission gases over catalysts – may couple phenomena from half-a-dozen disciplines producing a unique degree of complexity. Individual plasma effects combine *synergistically* across disciplinary domains to have unexpected consequences. Mastering this complexity to achieve the goals of an *FBRE* will require new convergent approaches that horizontally integrate disciplines.

At a high level, one could argue that the focus areas can be individually supported by different NSF programs, which is the present model. However, an LTP-focused program empowered to pursue common research challenges, catalyze convergent research and support the fundamental science of LTPs would enable advances towards the *FBRE* unachievable by other means. An LTP program would provide opportunities for future generations of scientists and engineers to pursue careers in which plasma science and engineering are the basis for providing societal benefit.

In Chapter III, the scientific background of LTPs will be discussed. The research challenges that have emerged in the four focus areas are discussed in Chapter IV. In some cases, the specific research challenges are closely tied to the applications and the ultimate goal of a *FBRE*. However, in each case there are over-arching, common research challenges that unify the field. Our concluding remarks, looking ahead, are in Chapter V.

## III. Scientific Background – The Science of Non-Equilibrium Plasmas Providing Societal Benefit

Low temperature plasmas are unique non-equilibrium systems. They typically have an electron temperature, $T_e$, which is much higher than the ion temperature, $T_i$, which is in turn often not much higher than the gas temperature, $T_g$. The ion and gas temperatures are generally close to ambient room temperature. Due to the partially ionized nature of LTPs, the plasma's specific energy content is low because the energy content is dominated by the far more abundant neutral gas. This situation provides a unique set of conditions wherein plasma species can non-destructively and beneficially contact surfaces. For example, the entire microelectronics industry, which forms the technological base of modern society, is enabled by the beneficial plasma-surface interactions which deposit and remove materials with nanometer spatial resolution in the fabrication of microprocessors [5]. This beneficial contact with surfaces now extends to liquids, such as plasma-activated water, which in turn has led to the emerging field of plasma medicine [6].

The fundamental LTP science issues revolve around controlling the distribution of energetic particles – most often electrons and ions. LTPs interact with atoms and molecules to produce excited states, radicals and photons. These species interact with surfaces for the purpose of beneficially modifying their properties and interact with dust or liquid aerosol particles in multi-phase plasmas. These interactions ultimately depend on the shape and evolution of the charged particle (electron, positive ion and negative ion) velocity distributions, $f(\vec{r},\vec{v},t)$. In fact, the ability to predict, control and shape $f(\vec{r},\vec{v},t)$ for beneficial interactions with atoms, molecules, solid- and liquid-phases is at the heart of advancing LTP science. Obtaining this predictive control is an incredibly challenging goal, *a grand challenge,* considering the extreme diversity and complexity of the field.

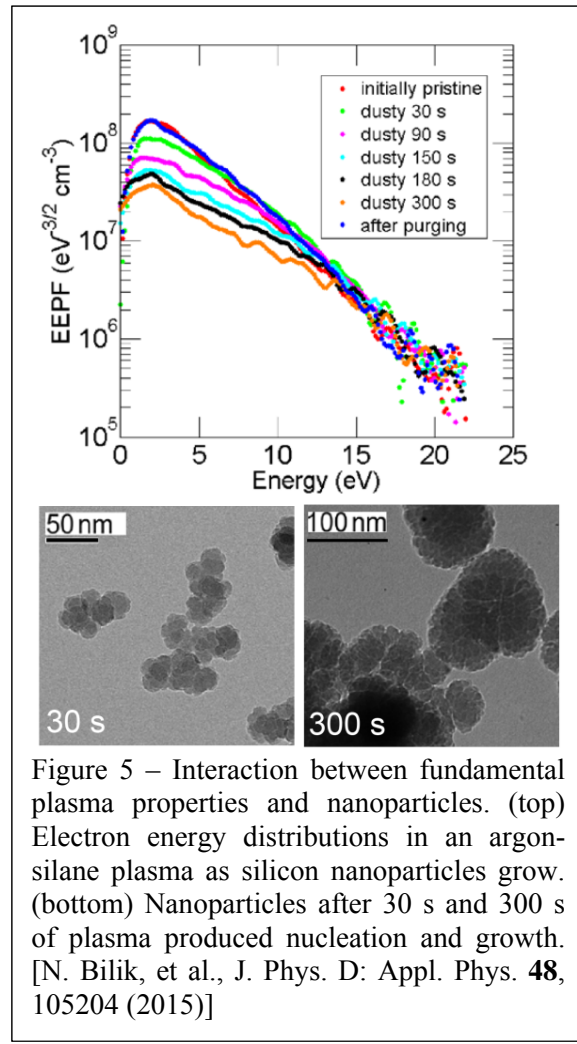

Figure 5 – Interaction between fundamental plasma properties and nanoparticles. (top) Electron energy distributions in an argon-silane plasma as silicon nanoparticles grow. (bottom) Nanoparticles after 30 s and 300 s of plasma produced nucleation and growth. [N. Bilik, et al., J. Phys. D: Appl. Phys. **48**, 105204 (2015)]

The relationship between $f(\vec{r},\vec{v},t)$ and the chemical reactivity it produces is clearly two-way and often very subtle. Consider multiphase plasmas, such as those used to produce nanoparticles for solar cells. The electron energy distribution (EED), derived from $f(\vec{r},\vec{v},t)$, is responsible for producing the chemical reactants that nucleate into the nanoparticles. The ion energy distributions (IEDs) striking the nanoparticles determine the crystallinity of the particles. The size and electrical properties of the nanoparticles then feedback to alter the plasma, perturbing and shaping the EED [14]. (See Fig. 5.)

One example of the complex chemistry resulting from electron impact reactions in atmospheric pressure plasmas comes from rare-gas plasma jets emitted into air for treating tissue in biomedical applications. Here, reactions are initiated by electron impact excitation and ionization of the argon and humid air impurities in the plasma jet:

$$e + Ar \rightarrow Ar^+ + e + e, \qquad e + H_2O \rightarrow OH + H + e, \qquad e + O_2 \rightarrow O + O + e.$$

These ions and radicals then initiate a cascade of complex chemistry that produces reactive species such as $HO_2$, $HNO_x$, $O_2(^1\Delta)$ and cluster ions such as $H_7O_3^+$. A state-of-the-art reaction mechanism that fully describes the production of reactive species air plasmas may contain nearly 100 species and 2000 reactions [15]. (See Fig. 6.)

LTPs encompass an enormous dynamic range of operating conditions. Typical areas being investigated by the LTP community span a range of $10^9$ in pressure (< 1 mTorr, in plasma etching, to liquid densities used in environmental applications and healthcare), $10^9$ in spatial scale (nanometers associated with plasma transport in nanoporous material, to meters in flat panel display film deposition) and $10^{12}$ in time (tens of picoseconds for formation of space charge layers in streamers to minutes in plasma surface interactions). The vast array of plasma chemical systems ranges from rare gases used in lighting to the multi-component gas mixtures employed in microelectronics processing (e.g., $Ar/C_4F_8/O_2/CO_2/N_2$). The bounding surfaces of these plasmas vary from polymers, metals, catalysts and semiconductors, to living tissue. This dynamic range of scientific investigation and applications, likely unique across engineering and the physical sciences, speaks to the convergent nature of the discipline.

Due to the large dynamic range of LTPs, there is no single overriding scientific challenge, beyond perhaps understanding the complex inter-linked processes that enable control of $f(\vec{r}, \vec{v}, t)$, that unites the field. There are however, highly linked and intermeshing sets of scientific and technological challenges that provide a broad front with which LTP science and technology frontiers can be advanced.

Although considerable scientific and engineering challenges face the LTP community, LTPs have already delivered enormous benefit to society. The following list of examples of societal benefit delivered by LTPs is incomplete but representative.

- The entire current and future information technology infrastructure owes its very existence to LTPs through their role in microelectronics fabrication [5].
- In 2012, 12% of the electricity generated in the US was expended by lighting and about 2/3 of that was used in LTP lighting sources [16].
- Renewable energy sources such as solar cell arrays, cannot be economically produced

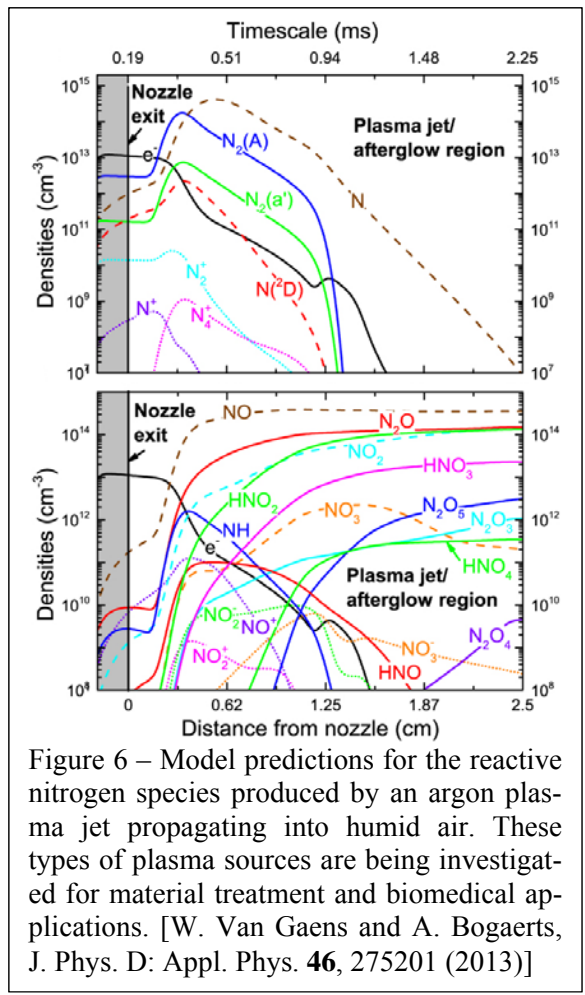

Figure 6 – Model predictions for the reactive nitrogen species produced by an argon plasma jet propagating into humid air. These types of plasma sources are being investigated for material treatment and biomedical applications. [W. Van Gaens and A. Bogaerts, J. Phys. D: Appl. Phys. **46**, 275201 (2013)]

- without thin film deposition and etching by LTPs [17].
- High efficiency jet engines would not exist in the absence of thermal barrier coatings produced by LTPs [18].
- Spacecraft rely on propulsion from LTP thrusters [19].
- A vast array of other technologies also would not exist, at least economically, without LTPs, including liquid crystal display (LCD) panels, mass produced polymer sheets, IR-filtering glazing on windows, hardened metals for human implants, industry pollution abatement devices, light emitting diodes (LEDs) [20] and high power lasers.
- Arrays of microplasmas are now used for sterilization and disinfection.
- The biotechnology and tissue engineering disciplines rely on LTPs for producing biocompatible surfaces [21].
- The thermal plasma spray global market alone is estimated to reach $12 billion by 2021 [22][23].
- There exists an emerging field of medical therapeutics in which atmospheric pressure plasmas are applied directly to human tissue for wound healing, infected tissue treatment and cancer treatment – plasma medicine [6].

Clearly, modern society would not be as advanced absent LTPs: imagine what *high technology* would mean if microelectronics were limited to early 1980s technologies, jet engines had not advanced since the days of the Boeing 707, and advanced human prostheses and implants were still objects of research. As evidenced by the broad range of applications above, LTPs have had and continue to have deep societal impact.

In addition to societal benefits, the field of LTPs has also made significant contributions to plasma science: LTPs have been the source of many fundamental physical principles that form the basis of other fields of plasma physics. Primary concepts of electron and ion transport, cyclotron resonance, electromagnetic wave interactions with plasmas, electrical probes, interferometric diagnostics, charged particle distribution functions, high energy beam produced plasmas, laser diagnostics, radiation transport in plasmas and non-ideal plasmas were all first developed (and continue to be developed) in the context of LTPs. The field of LTPs continues to hold extreme scientific challenges, largely centered on the control of power through the plasma for the selective production of excited states, ions, photons, and surface reactivity.

## IV. Science Challenges in the Focus Application Areas

The unifying high-level science challenges discussed in Chapter II extend across the four focus application areas. To achieve the goals of an *FBRE*, the science challenges must closely link to the intended applications. This chapter discusses science challenges motivated by the four focus application areas. Many of these science challenges are common to multiple focus areas, such as the challenges in modeling, simulation and diagnostics. These overlapping challenges will be discussed in more detail in the first section on multiphase plasmas.

### *IV.A. Multiphase Plasma Systems*

#### IV.A.1. Background

A multiphase-plasma (MPP) is a partially ionized gas in which solid or liquid materials are dispersed within the gas phase, as in an aerosol-laden or dusty plasma, or where reactivity is transferred between phases, as in a gas phase plasma interacting with a liquid. Under these conditions, the characteristic length and time scales governing the MPP's behavior are inconsistent with the scaling that governs either phase individually. For example, dusty plasmas have solid particles suspended in the plasma that are much smaller than the Debye length that defines the scale over which collective plasma behavior may occur [24][25]. Likewise, plasma interaction with liquids can produce electric fields and chemical gradients in the liquid phase whose dimensions approach the nanoscale, potentially requiring atomistic level simulation to elucidate this multiphase interface [26][27].

These (usually) condensed phases interact with the gas phase to form a synergistic system whose description requires a self-consistent treatment of the gas plasma, the condensed phase, and the interfacial transport between them. This description must span lengths from nanometers to centimeters and time scales from nanoseconds to days. Examples of MPPs include dusty plasmas (gas phase plasmas interspersed with solid particles) [28][29][30][31][32], plasmas in or in contact with liquids, laser ablation plumes [33], plasmas in or in contact with porous materials and plasma in bubbles in liquids. An extreme case of an MPP is a vacuum arc where the gaseous plasma is produced from evaporated electrode material and injected into a vacuum. These highly coupled systems with poorly understood interfacial conditions require investigations that go well beyond the traditional approach in which plasma-surface interactions are considered one-way: plasma to surface. That is, the flux of reactant species from the plasma affects the surface, but the surface only nominally affects the plasma. MPPs invariably have synergistic two-way coupled interactions across interfaces between the plasma and a complex condensed-phase medium. The coupled interaction of the plasma with the condensed phase considerably changes both the condensed phase and the gas plasma [34][35][36]. (See Fig. 7.)

Plasma phenomena at liquid-vapor interfaces are particularly critical multiphase systems due to their compelling yet minimally understood science and their significant impact across a broad range of applications providing substantial societal benefit. While these systems drive environmental applications, water purification, fuel reforming, plasma aided combustion and biotechnology, they are poorly understood [37]. These interactions include gas-phase plasmas intersecting with a liquid (but not sustained inside the liquid), plasmas inside macroscopic bubbles in the bulk of the liquid, plasma-enveloping aerosols in the gas phase and plasmas fully sustained in the liquid phase [38]. The synergistic processes in these interactive systems are not fully captured by any single-phase or adjacent-phase model [39]. Due to the strong exchange of matter, charge and energy across a dynamic interface, decoupled models for individual phases are not able to cap-ture the plasma kinetics, interfacial and interphase phenomena and phase transitions with other materials. The lack of self-consistent physical models indicates there are fundamental issues in MPP still needing resolution. For example, researchers debate whether plasma formation in liquids precedes formation of gas micro-bubbles or vice-versa. A self-consistent physical model coupling transport and transformation of charged and neutral chemical species and energy in all forms and phases is needed to better understand MPP involving liquids as compared to simpler (one-phase) gaseous or plasma-solid interactions.

The thermal spraying community has investigated the transport of droplets containing solutions for coatings production. Although these systems are often in quasi-equilibrium, they share many of the same science challenges discussed here for non-equilibrium systems. For example, a key challenge in thermal spray coating is controlling the coupling between a dispersed liquid phase and the plasma [40].

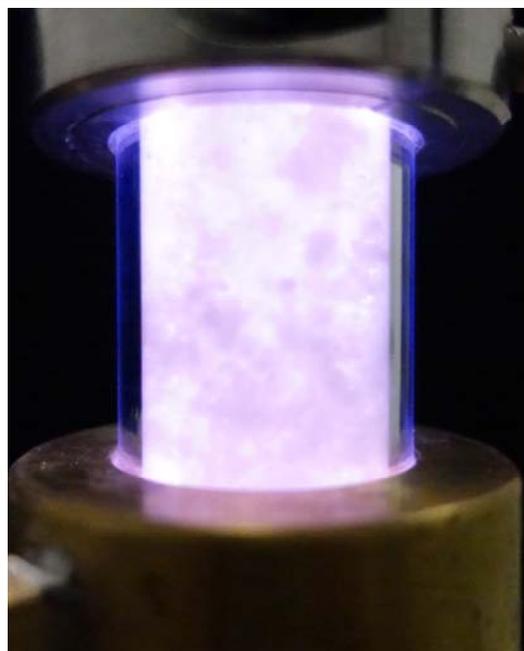

Figure 7 – Advanced plasma-materials interactions in multiphase systems will enable new regimes of chemical processing, from production of nanoparticles to the plasma enhanced fluidized bed reactor pictured here. [E. Thimsen, private communication, (2016)]

To provide a context for this complexity, consider the interaction of an atmospheric plasma jet with a liquid, as might occur in the treatment of a biological system [41]. A rare gas seeded with a small percentage of a reactive gas (e.g., $He/O_2$ = 99/1) flows through a cylindrical plasma tube and mixes with room air. A plasma ionization wave (IW) is generated in the tube, propagates at 100s km/s through the plume and contacts the liquid. During its propagation, the IW produces reactive species as the ambient air diffuses into the plume. The IW may trigger turbulence in the plume by electrical and physical forces [42]. The IW penetrates through a layer of the saturated liquid evaporating from the surface. Upon striking the liquid, the IW wave propagates across the surface of liquid, in some cases producing self-organized patterns. (See Fig. 8.) Reactive gas phase species solvate at and transport across the gas-liquid interface. This initiates a cascade of reactions in the liquid, while photons and electric fields from the plasma transport energy into the liquid. The properties of the liquid (e.g., electrical conductivity and/or chemical composition) in turn alter the plasma both physically and chemically. The plasma interaction at the interface induces convection in the liquid [43]. The transformation of plasma produced active species into liquid residence activation occurs on spatial scales of nm to mm. Recent findings suggest that the interfacial region between the plasma and the solution can play a key role in this transfer, particularly for processes involving highly reactive short-lived species [26][27][44]. In cases of living systems, treating cell cultures or human tissue for wound healing, there is a biological response to these plasma activated processes.

Important processes in this specific example include: (i) propagation of surface ionization waves, plasma self-organization, and instability development over a liquid-vapor interface; (ii) surface charge accumulation, solvation and transport and associated effects on the plasma-liquid interface; (iii) coupled radical kinetics transport in the vapor phase; (iv) ionization and charge transport in dense media; (v) ion-molecule chemical reactions and multi-body collision-induced processes in the liquid phase; (vi) effects of these collisions on liquid-vapor phase equilibrium; (vii) extremely high spatial gradients and short relaxation lengths and quenching times; and (viii) biological responses to these processes. These processes demonstrate the extreme complexity and diversity of the LTP discipline.

By definition, MPPs are in extreme thermal non-equilibrium, which results in gaseous species having different translational and internal (rotational/vibrational) temperatures and electrons not having well-defined temperatures at all. Moreover, the interspersed liquid or solid phases may have temperatures that differ from the temperatures of the gaseous species. For instance, nanoparticles in a plasma may have temperatures greatly exceeding the gas temperature due to non-equilibrated plasma-nanoparticle energy exchanges [ 45 ]. These abnormally high particle temperatures enable crystallization of the nanoparticles. (Nanoparticle-plasma interactions are discussed in Sec. IV.D in the context of materials fabrication.) Conversely, liquid droplets immersed in plasmas may have temperatures much lower than the surrounding gaseous species through plasma-enhanced evaporative cooling.

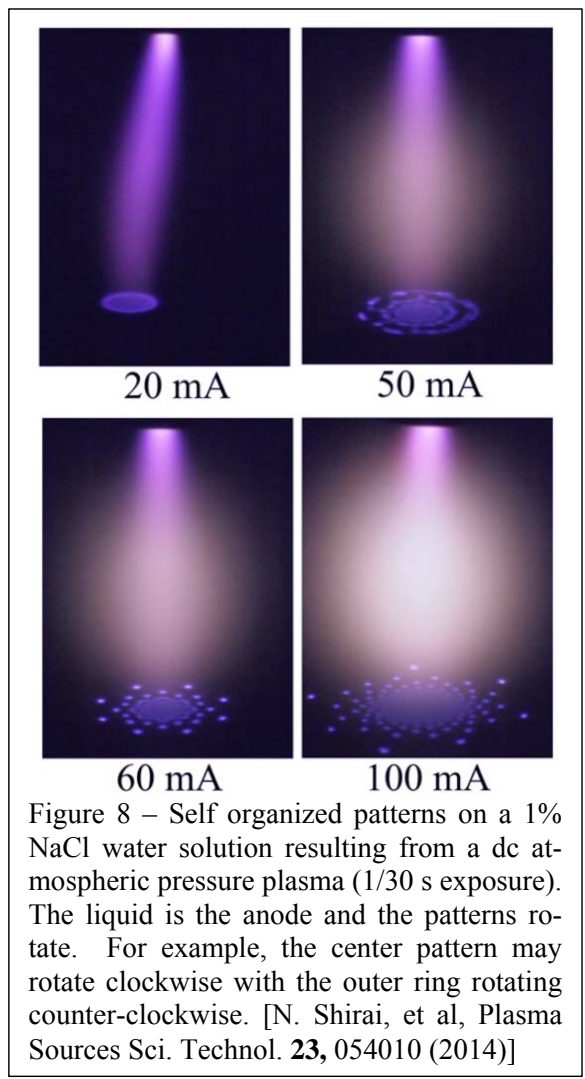

Figure 8 – Self organized patterns on a 1% NaCl water solution resulting from a dc atmospheric pressure plasma (1/30 s exposure). The liquid is the anode and the patterns rotate. For example, the center pattern may rotate clockwise with the outer ring rotating counter-clockwise. [N. Shirai, et al, Plasma Sources Sci. Technol. **23,** 054010 (2014)]

IV.A.2.    Synergies, Opportunities, Challenges

The implications of MPP on the *FBRE* and the *food, energy, water nexus* are profound. MPPs are utilized in water purification, fuel reforming, production of advanced materials for photovoltaics, catalyst enabled selectivity and attacking antimicrobial resistance (AMR). The common scientific challenges of MPPs can be dominated by controlling the flow of power through the plasma to selectively activate desired outcomes in one or both phases. Depending on the application, these desired outcomes may be very different: efficiency, uniformity or process stability are some examples. In biological applications, control and reproducibility are of paramount importance and energy efficiency is typically of secondary importance. In contrast, for large scale energy applications, plasma chemical processes may require minimizing the specific energy cost of converting one material into another. The inherent complexities of MPPs make designing and controlling such systems extremely challenging. Developing expertise in the associated chemical, material and biological sciences, coupled with MPP, demands convergent research.

If the science and technology of MPPs follow past successful applications of plasmas to societally significant problems, MPP chemical reactors will likely utilize high pressure systems to maximize throughput. Such systems will likely be relatively high power density and highly collisional, potentially leading to both plasma and thermal instabilities, which may also destabilize the phase boundaries. Controlling these instabilities requires a fundamental understanding of the stability criteria for both phases and insight into how those phases may interact under the electrical and other forces acting on the plasma. There is currently no approach based on fundamental understanding of the interfacial transport to achieve this stability through feedback control or other means. Adding another layer of complexity, timescales associated with control in the two phases can differ by many orders of magnitude.

Fundamental investigations of MPPs will require significant advances in computational techniques. The computational challenges associated with MPPs can be divided into at least five categories: (i) high spatial resolution is required in critical portions of the system (ionization waves, sheaths, material interfaces, double layers within the liquid); (ii) kinetic descriptions of transport may be required even at high pressures if there are sharp gradients; (iii) disparate time-scales for electron, ion and neutral transport, particularly in different phases, require long simulation times if all timescales are simultaneously resolved, which in most cases is not practical; (iv) coupling electromagnetics with charge transport results in highly non-linear problems (especially when ionization processes are important); and (v) interfacial phenomena and phase transitions must be included. The first four challenges are relevant to all LTPs and the four focus areas, the last is specific to MPP.

Although there has been impressive progress in modeling MPPs [46][36], current computational tools are challenged to fully capture the known physics in a unified approach. Computational models for many MPP problems cannot simulate the coupled physics and chemistry on a timescale useful for practical applications. For example, simulations of the ionization waves that occur in most atmospheric pressure plasmas require high spatial and temporal resolution of dynamically evolving shock-like fronts. When these ionization fronts interact with liquids such as water, they generate chemical and electric field gradients that approach the intermolecular spacing of the liquid, potentially requiring an atomistic approach such as molecular dynamic simulation. This is impractical with the static (non-moving) meshes used in most existing codes. Adaptive, multiscale kinetic-fluid solvers, using adaptive mesh and algorithm refinement, could in principle address these challenges if adapted to MPPs [47][48]. Resolving challenges associated with disparate time and length scales and the non-linear nature of plasma equations requires novel methods and algorithms [49].

An impressive set of gas phase diagnostics is available to measure many plasma parameters, including species concentrations, temperatures and internal electric fields. (See Fig. 9.) As with many diagnostics, each technique has limitations imposed by conditions of the system itself, as well as the practicalities of physical, optical or electrical access to the system. For example, the Langmuir probe is a diagnostic tool commonly used in plasma science. However, these probes are difficult to use in MPP systems due to higher gas pressures, reduced scale lengths and limited physical access to regions of interest in the plasma. Addressing the scientific challenges of leveraging LTPs for sustainability requires measuring increasingly complex non-equilibrium chemistries and complex molecules in systems with extraordinarily small scale lengths and times, additionally challenged by turbulence and stochastic processes. Optical and in-situ electrical measurement techniques provide the most direct path to measuring MPP systems, but the following challenges illustrate the complexities inherent with MPPs.

- Reliable measurements in high pressure systems will demand picosecond and nanosecond fluorescence techniques and absorption-based diagnostics.

- Interpreting plasma optical emission spectra remains a challenge due to complex kinetics, high collisionality and lack of relevant cross sections required for collisional-radiative models. The stochastic character of MPPs complicates interpretation of diagnostics.
- Advanced optical diagnostics for MPPs are practically unexplored because of limited optical access as well as screening and scattering by droplets and aerosols.
- Development of novel laser diagnostic techniques that can distinguish between phases and be applied in single pulse mode is critically needed.

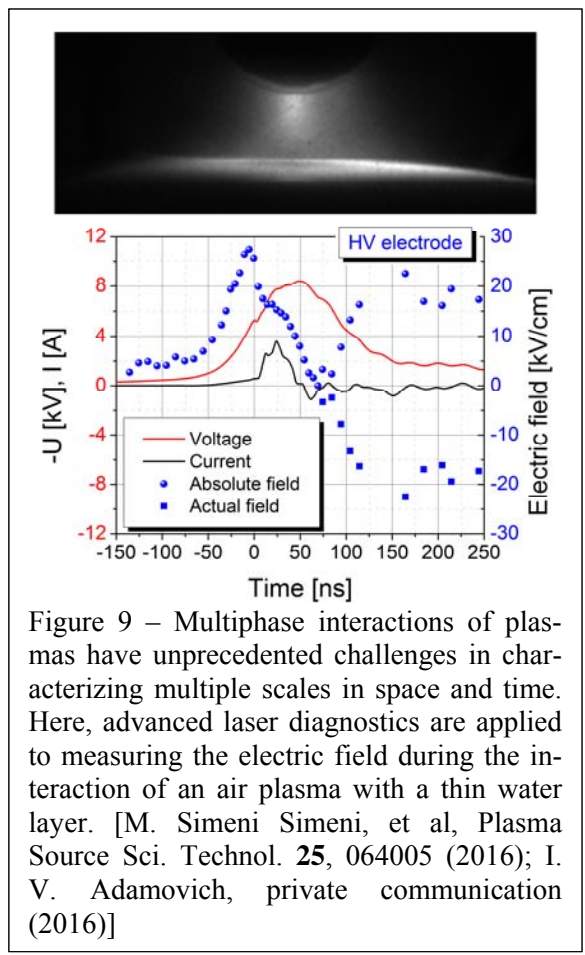

Figure 9 – Multiphase interactions of plasmas have unprecedented challenges in characterizing multiple scales in space and time. Here, advanced laser diagnostics are applied to measuring the electric field during the interaction of an air plasma with a thin water layer. [M. Simeni Simeni, et al, Plasma Source Sci. Technol. **25**, 064005 (2016); I. V. Adamovich, private communication (2016)]

The transfer of plasma produced reactivity to a liquid is a complex process involving transport and interfacial phenomena. There currently does not exist a suite of measurement techniques that can quantify short-lived reactive species in-situ in both phases with sufficient sensitivity *without perturbing the system* to quantify reaction mechanisms, and thus enable real-time control.

Underlying MPP science discussions is the issue of reaction mechanisms. As discussed in the recent review, "Plasma–Liquid Interactions: A Review and Roadmap" [37], the physical mechanisms underlying MPPs, particularly plasma-liquid interactions, are poorly understood at best. Currently, no model can reliably predict from first principles how chemically reactive plasmas couple with bounding liquids. For example, reaction mechanisms for air plasmas activating pure water are only now beginning to approach predictability [50]. However, water with only mild organic contamination is beyond our ability to quantify plasma-produced reaction chemistry. The situation with biological liquids or water with complex contamination is even more daunting.

IV.A.3.    Summary of Science Challenges

The science challenges of MPPs are extraordinarily broad due to the extreme diversity of the field. The following goals associated with MPPs were identified at the workshop:

- *Controlling the flow of power through the plasma to selectively activate the desired outcome in one or both phases* – An example of this scientific challenge is plasmas having internal phase boundaries, such as nanodusty plasmas or plasmas laden with aerosol droplets. Reactive species must be selectively produced to enhance the plasma-chemical interaction with these internal phases, while avoiding possible plasma-induced damage to the nano- and micro-structures.
- *Understanding self-organization and controlling instabilities* – It is difficult to understand self-organization and control instabilities even in single-phase plasmas. The inherently non-linear feedback between phases in MPPs, which are often strongly coupled systems, adds complexity and scientific challenges. Without controlling these nonlinear features, efficient, targeted, reproducible chemical activation in MPPs will not be achieved.

- *Developing multiphase models and diagnostics* – The increased complexity of models and diagnostics in multiphase systems results from interfacial transport which often manifests itself over small spatial scales but has global impacts. Current modeling and diagnostic techniques fail to address these complexities. Increased fidelity in scale (length and time) and extension to atomistic approaches in simulation are needed. Similarly, these length and time scales demand new methods of measurement that can probe across similar ranges of time and space, as well as function in the typically harsh chemical and physical environments experienced in chemically active LTPs and their adjacent phases.
- *Developing reaction mechanisms* – Mechanisms associated with plasma-solid surface interactions continue to challenge plasma science but are relatively simple compared to those associated with plasma-liquid interactions. In the latter case, not only does vapor-liquid interfacial transport complicate the problem, but interfacial electrochemical phenomena, liquid surface instabilities and sub-surface liquid convection make the problem more complex than the corresponding plasma-solid surface interactions. Developing multiphase plasma initiated reaction mechanisms that account for this complexity is critical to not only MPP but to all focus areas.

## *IV.B.    Energy and the Environment*

### IV.B.1.    Introduction

The use of LTPs in the areas of energy and the environment hold perhaps the greatest potential for applications leading to a more sustainable future. Plasmas are capable of initiating processes with unique selectivity and unmatched energy efficiency, due to their ability to operate in a state of non-equilibrium. Plasma-initiated reactions can generate radicals, excited states and photons that are simply inaccessible to purely thermal, equilibrium systems – and this can be done at near ambient temperatures. Examples of selectively channeling non-equilibrium plasma power into preferred states of atoms and bonds of molecules include atomic, molecular and excimer based lighting sources and lasers, ozone production, materials processing for photovoltaic cells and microelectronics, and high power switches. The key challenge is extending this record of selectivity and efficiency to emerging applications connected to energy and the environment.

Plasmas improve sustainability by replacing energy-intensive near-equilibrium (thermal) processes with more energy efficient non-equilibrium processes. Of all the environmental issues that must be addressed, $CO_2$ engineering is perhaps the most pressing. There is currently no technology available to economically and permanently remove $CO_2$ from the environment, or to capture and recirculate the carbon in a carbon neutral manner. For example, research is addressing plasma conversion of $CO_2$ to CO for syngas (a mixture of CO and $H_2$) to recirculate the carbon for carbon-neutral combustion [51][52]. Doing so with plasmas is exceedingly challenging since power is needed to generate the plasma and that power expenditure should itself not contribute to $CO_2$ generation. Despite impressive progress, plasma based conversion of $CO_2$ is not viable unless the power is generated from purely green and renewable sources. Similar LTP processes are being investigated for plasma conversion of $CH_4$ to hydrogen (the second component of syngas) [53] and to higher value $C_xH_y$ hydrocarbons [54].

IV.B.2. Synergies, Opportunities, Challenges

Utilizing plasmas in energy and environmental applications already has a record of success. Thin film solar cells are economically viable due to the efficiency and selectivity of plasma-assisted deposition and thin-film etching in industrial scale fabrication processes [17]. Plasmas are essential to high efficiency lighting and thin film light emitting diode (LED) fabrication [20]. $NO_x$ remediation technology has been successfully developed using plasmas [55][56]. Pilot plants use plasma torches for converting municipal solid waste (MSW) to syngas and minimizing the need to dispose of solids [57]. Other examples of current investigations of LTPs in energy and environmental processing and their challenges include:

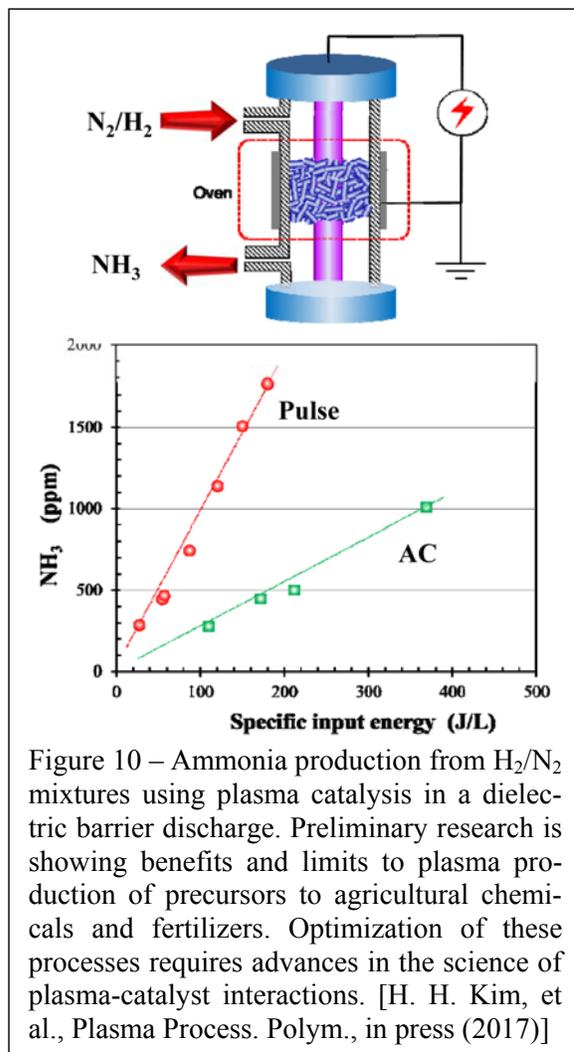

Figure 10 – Ammonia production from $H_2/N_2$ mixtures using plasma catalysis in a dielectric barrier discharge. Preliminary research is showing benefits and limits to plasma production of precursors to agricultural chemicals and fertilizers. Optimization of these processes requires advances in the science of plasma-catalyst interactions. [H. H. Kim, et al., Plasma Process. Polym., in press (2017)]

- *Nitrogen fixation*: The worldwide agriculture infrastructure relies on nitrogen fixation for fertilizers. Arguably our worldwide food cycle is overly dependent on the Haber-Bosch process for nitrogen fixation, a process that has reached its efficiency limits and is not further scalable upwards to larger rates of production or downwards to highly energy efficient point-of-use fertilizer production. LTPs are already showing promise in providing nitrogen based plant nutrients through non-equilibrium plasma processes. This is discussed further in Section IV.C. on agricultural applications of LTPs. (See Fig. 10.)
- *Carbon dioxide conversion*: Currently, emission of harmful gases, such as $CO_2$, into the atmosphere is perhaps the single greatest environmental challenge. The first line of defense is reducing emission of these gases. The second line of defense is capturing and either sequestering or converting $CO_2$ into useful products. Intensive, multidisciplinary research is now addressing the plasma based conversion of $CO_2$ into industrially usable chemicals.
- *Methane activation and conversion:* There will always be a demand for complex carbon based molecules for materials and chemicals. Efficiently converting abundant methane into these complex carbon molecules using selective plasma chemistry is a major challenge but offers promising advantages over conventional methods.
- *Water purification and reclamation:* Many municipal water systems already rely on plasmas for purifying water – plasmas produce the ozone that is used for water purification worldwide. However, challenges remain. Conventional methods struggle with resistant pollutants (from pharmaceuticals to viruses) which are endemic in global water systems. Better point-of-use water purification methods, either for human consumption or at the output of industrial processes, are required. Plasma based water purification and reclamation, through ad-vanced oxidation technologies, represent a tremendous opportunity.

- *Pollutant mitigation and waste treatment*: Although the goal is to have 100% recyclable materials and processes, it is likely that there will always be a waste stream that must be treated before being released to the environment. Plasma based pollution mitigation and waste treatment is already making such inroads. Plasma based systems are used to treat contaminants in industrial gases, to treat $SO_x/NO_x$ emission from power plants, to remediate medical waste and to convert municipal waste into syngas for sustainable power production.
- *Combustion enhancement*: Fossil fuel combustion will continue to play a vital role in modern society for the foreseeable future. Using those precious resources more efficiently, positively impacts every measure of environmental stewardship. Plasma aided ignition and combustion (PAIC) is a highly promising field of research.

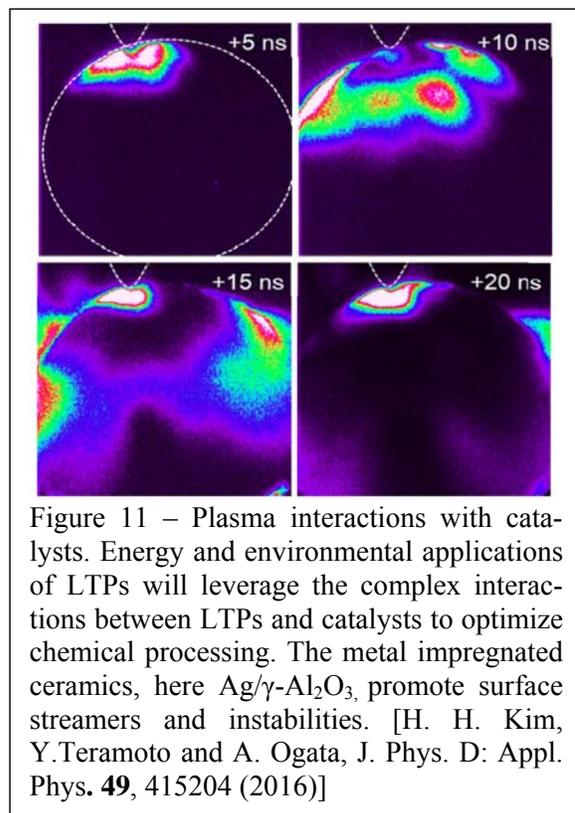

Figure 11 – Plasma interactions with catalysts. Energy and environmental applications of LTPs will leverage the complex interactions between LTPs and catalysts to optimize chemical processing. The metal impregnated ceramics, here Ag/γ-$Al_2O_3$, promote surface streamers and instabilities. [H. H. Kim, Y.Teramoto and A. Ogata, J. Phys. D: Appl. Phys. **49**, 415204 (2016)]

- *Bio-based carbon conversion:* The ultimate closed carbon cycle uses bio-based carbon to produce the chemicals societies require. Recent research indicates LTPs can play a key role in this effort.

Plasma-assisted catalytic conversion of complex molecules is one of the frontier areas of research with perhaps the greatest potential impact on sustainability [58][59]. Conventional catalysis is one of the most fundamental and critical processes in the chemical industry. Catalysts enable reactions to occur at higher rates at lower temperatures and in a more selective fashion. Plasma catalysis utilizes the ability of LTPs to produce excited states (vibrational and electronic) and radicals through electron impact dissociation to more selectively utilize catalysts at lower temperatures. Plasma catalysis has been investigated for $CO_2$ and $CH_4$ conversion to produce more complex higher value chemicals [53][54], and for remediating environmental toxins [60]. In contrast to conventional thermal catalysis, low temperature plasma activated catalysis has the potential to enable new chemical processes. For example, production of thermally unstable chemicals can be realized with the LTP activation of catalysis. In principle, this process is fully sustainable. The plasma portion of the chemical conversion can be powered fully on renewable sources of electricity. The chemical feedstock can, in principle, be from any bio-generated hydrocarbon. Plasma chemical processes, powered sustainably by renewable energy, could generate a suite of products including synthetic fuels, platform chemicals and complex molecules with little environmental impact. One key challenge is the plasma-catalyst-surface interaction. (See Fig. 11.) No existing fundamental theory provides guidance on optimizing this interaction for a specific chemical system [58]. Investigations have been dominantly empirical. This is another example of convergent research – the complexity of understanding and leveraging the synergistic interactions between plasmas and catalysis will not be mastered by a single discipline.

Although glow discharge plasmas are often thought of as being destructive through dissociation reactions, they can be effective in synthesizing and chemically modifying organic compounds. Example reactions include polymerization, isomerization, elimination, rearrangement, decarbonylation, and oxidation, including both alkane and aromatic compounds [61]. The majority of these reactions are highly selective and result in products of high yields and efficiencies, thereby offering interesting synthetic possibilities.

While non-equilibrium plasmas have been the focus of this discussion, thermal plasmas are exceedingly important in environmental and energy stewardship. (Thermal plasmas are discussed in more detail in Section IV.D.) For example, environmental concerns regarding chrome electroplating using hexavalent chromium have led to the replacement of hard chrome coatings by the "greener" thermal plasma spray coatings. Utilizing thermal plasmas for biomass gasification and conversion of municipal solid waste (MSW) to syngas is expected to be extremely important to a carbon-neutral economy [57]. Environmental and energy applications of thermal plasmas inevitably result in a unique mix of thermal and non-thermal regions of the plasma, introducing exceedingly sharp spatial gradients in both temperature and material properties. This mix of thermal and non-equilibrium conditions includes all the challenges that strictly non-equilibrium plasmas processes have – and perhaps more.

For most target applications, process selectivity, conversion, energy efficiency and scale-up are still major challenges. Considering LTP processes will likely be heterogeneous systems, addressing these challenges requires developing an understanding of the fundamental plasma-surface (interface) interactions in an integrated approach that combines new computational strategies with diagnostic techniques. Although there are guiding scientific principles that address all such systems, many of these basic studies will need to be performed in the context of specific applications – the complexity and diversity of these systems makes a single scientific solution difficult to universally apply.

For example, consider heterogeneous plasma-catalysis systems for $NO_x$ reduction, $CO_2$ conversion or fuel reforming. The chemical pathways for $NO_x$ reduction and fuel reforming are likely based on radical chemistry. The chemical pathways for $CO_2$ conversion are likely vibrationally enhanced thermal dissociation. The fundamental reaction pathways, plasma reactors, controls systems and process flow for these two approaches are very different, which makes a single universal science issue difficult to isolate. Having said that, there are likely more widely applicable principles that may be investigated in the synergetic effect of combining the plasma with a catalyst simply because so little is known of these processes. Fundamental studies will include precise measurements and modelling of excited, ionized species and radicals in the plasma, and the nature of the species interaction with the surface of the catalyst (e.g., species penetration depth into pores).

Electrical discharge plasmas can be credited with the development of the first Advanced Oxidation Process (AOP) for water treatment: ozonation. However, since its first use early in the $20^{th}$ century, ozone has been replaced with chlorine and AOPs such as ozone-hydrogen peroxide and ultraviolet light-hydrogen peroxide. Compared to these conventional AOPs, electrical discharge plasmas have several advantages; they require no chemical additions, degrade a broader range of contaminants and have the potential to be optimized for small treatment systems. Plasmas are capable of generating all the chemical species and effects found in the other AOPs as well as additional factors not typically found in those processes. These include the reactive oxygen species (ROS), (e.g., •OH, $H_2O_2$, $O_3$), (V)UV emission, and shockwave formation (when the discharge occurs directly in the liquid). The abundance of these factors and the efficiencies of their generation vary with the plasma reactor design and the type of plasma. In addition to the generation of the ROS found in most AOPs, plasma processes can generate reactive nitrogen species (RNS) and chemically reducing species.

The primary challenges for plasmas employed in energy and environmental applications involving contact with liquid surfaces, such as plasma activated water and waste treatment, are centered around interfacial phenomena. These processes result in plasma produced radicals in the liquid, either beginning with a gas phase plasma or creating plasma in the liquid, directly or through bubbles. These challenges are discussed in Section II.A.

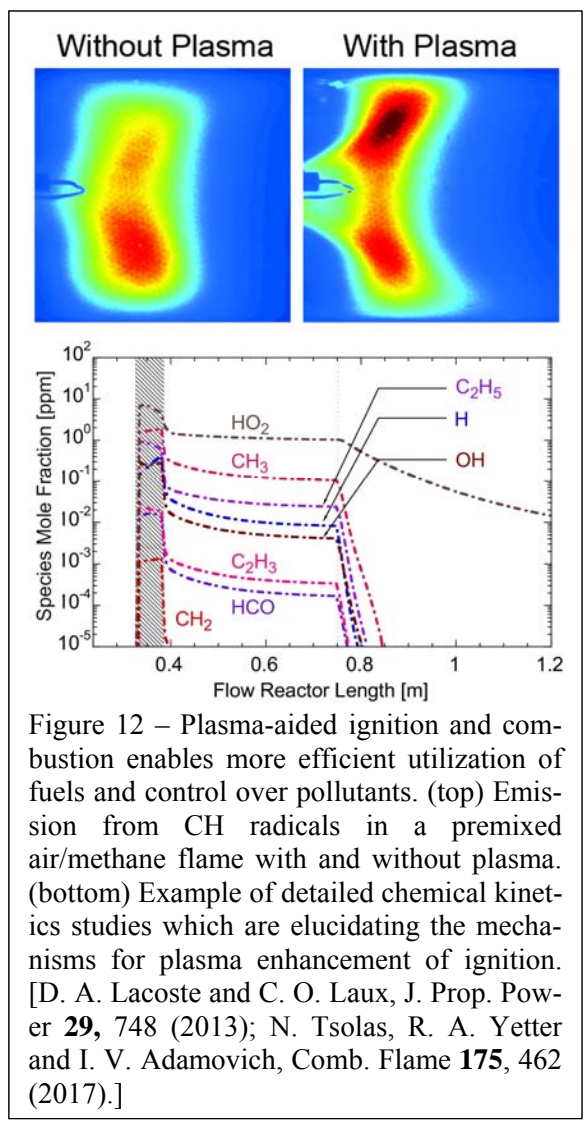

Figure 12 – Plasma-aided ignition and combustion enables more efficient utilization of fuels and control over pollutants. (top) Emission from CH radicals in a premixed air/methane flame with and without plasma. (bottom) Example of detailed chemical kinetics studies which are elucidating the mechanisms for plasma enhancement of ignition. [D. A. Lacoste and C. O. Laux, J. Prop. Power **29,** 748 (2013); N. Tsolas, R. A. Yetter and I. V. Adamovich, Comb. Flame **175**, 462 (2017).]

Fossil fuel combustion is expected to persist as society transitions to renewables and may continue far into the future for certain high value processes, such as aerospace applications. The key challenge is performing that combustion more efficiently in conditions that produce fewer pollutants. This is the goal of plasma aided ignition and combustion (PAIC) [62]. (See Fig. 12.) Promising results, for example extending the lean limit to reduce emissions of nitrogen oxides ($NO_x$), have been demonstrated with nanosecond-pulsed plasmas, as well as with microwave, laser and other sources. The primary challenges facing the field of PAIC include: (i) optimizing the energy partition in the plasma between electronic excitation of molecules/atoms including molecular dissociation, vibrational modes, and the consequences of prompt translational energy generated from dissociative excitation; (ii) developing an understanding of how excited electronic states of atoms, as well as ground electronic states of radicals, affect fuel-air plasma chemistry; (iii) determining the effect of vibrationally excited molecules on rates of plasma-chemical reactions; and (iv) developing plasma sources with the necessary reliability and cost for real world use.

Plasma actuators, using LTPs to improve the aeronautical efficiency of airflow over wings, is another example of plasmas impacting the energy efficiency of critical energy-consuming infrastructures. The initial investigations were intended to reduce drag and prevent separation of airflow over airplane wings [63]. Recent investigations have applied these flow-control principles to improving the efficiency of electrical power generation by wind-driven turbines [64].

IV.B.3.     Summary of Research Challenges

- *Understanding the flow of power through complex plasma chemical systems* – More than any other area of plasma physics and engineering, the investigation of LTPs for environmental stewardship will need to identify how power flows through a complex plasma chemical system bounded by complex materials, resulting in breaking perhaps a single chemical bond. In energy applications, efficiency at large scales is perhaps more important than in any other application of LTP. If plasma-assisted chemical activation could approach twice the efficiency of today's conventional technologies, arguably our impending environmental crises would be more manageable. (See Fig. 13.)
- *Understanding plasma-surface (multi-phase) interactions at the atomic level* – Determining the role of free electrons and large-electric fields on the chemical reactivity of surfaces will be necessary to optimize plasma catalysis and convert bio-feedstocks to chemicals, for example. Electrically charging even a non-catalytic surface affects rates of plasma-produced radical chemistry, and this important process is now virtually uncharacterized.
- *Investigating methods to design and optimize chemical processes* – Leveraging convergent research between the chemical physics and biotechnology disciplines to investigate inverse methods to design and optimize target chemical processes is a challenge that is virtually untouched.

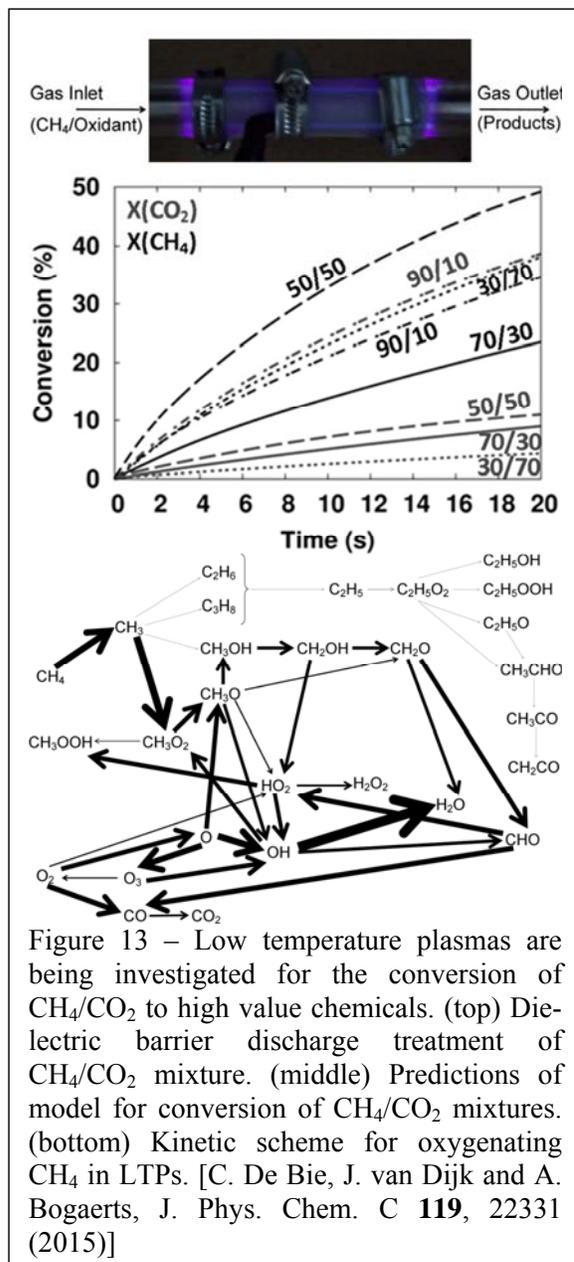

Figure 13 – Low temperature plasmas are being investigated for the conversion of $CH_4/CO_2$ to high value chemicals. (top) Dielectric barrier discharge treatment of $CH_4/CO_2$ mixture. (middle) Predictions of model for conversion of $CH_4/CO_2$ mixtures. (bottom) Kinetic scheme for oxygenating $CH_4$ in LTPs. [C. De Bie, J. van Dijk and A. Bogaerts, J. Phys. Chem. C **119**, 22331 (2015)]

- *Understanding and improving fundamental plasma processes* – Near-term benefits of plasmas for energy and the environment will likely come from improving existing processes, such as plasma aided ignition and combustion, and plasma enabled aeronautical flow control. The first challenge is understanding how fundamental plasma processes can be inserted into the huge in-place infrastructure for these technologies.
- *Identifying new computational and diagnostics methodologies* – All the challenges are multi-scale (in both space and time), from atomic-level to reactor-level, and from picoseconds to minutes. New computational and diagnostics methodologies are needed to simultaneously probe all relevant scales. The resulting insights will enable reduction of thousands of non-linearly coupled parameters to smaller, more manageable sets upon which engineering advances can be based.

## IV.C. Biotechnology and the Food Cycle

### IV.C.1. Introduction

LTPs have been incredibly impactful in biotechnology, medicine and the food cycle. Most artificial joints, stents and biocompatible implants, as well as large arrays of medical devices are fabricated using LTPs for surface hardening, depositing coatings and shaping materials [65][66]. An entirely new field of *plasma medicine* – the direct use of plasmas in human healthcare – has emerged in the last decade [67][68][69][70]. Even more recently, the field of plasma-agriculture has re-emerged in which plasmas are used to enhance agricultural production, from stimulating growth, treating waste and reducing contamination, to plasma based fertilizers [71]. (See Fig. 14.)

The advent of atmospheric pressure plasmas having ion and neutral temperatures close to room temperature (cold atmospheric plasma – CAP) has led to emerging applications in biomedicine that pose no risk of thermal damage to tissue [72][73]. CAP's unique chemical and physical properties have enabled a broad array of biomedical applications. The advantages of CAP are likely due to the production and efficient delivery of reactive oxygen and nitrogen species (RONS), analogous to those naturally produced by cells [74]. CAP is capable of gentle non-thermal modification of the radical balance in cells. A few successes in plasma-biomedicine follow.

*Cancer Treatment*: Recent research indicates CAP selectively eradicates brain tumor cancer cells in vitro without damaging normal cells and significantly reduces tumor size *in vivo*, which may lead to a new non-invasive surgery that allows specific cell removal without affecting the entire surrounding tissue [75]. In studies using mice models, *in vivo* tumor growth was significantly slowed or the tumor eradicated, and survival rates increased with CAP treatment on the skin of mice [76][77]. One promising development is that the CAP treatment activates the immune response *in vivo* to attack the tumor [78].

*Antimicrobial and Antiviral Properties*: CAP effectively treats bacteria in biofilms and on wound surfaces [79]. Through *in situ* production of RONS, CAP inactivates bacteria and viruses. A recent study concluded that treating cells with cold plasma leads to their regeneration and rejuvenation. (See Fig. 15.) From this result, a plasma therapy program for patients with non-healing wounds can be developed [80]. Studies of CAP on HIV-1 replication demonstrated that pre-treatment of infected cells with CAP inhibits virus-cell fusion, viral reverse transcription and integration. Virus particles produced by CAP-treated cells had reduced infectivity [81].

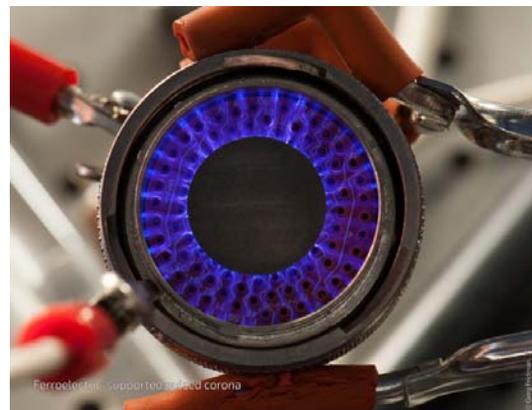

Figure 14 – Nanosecond-pulsed plasma treatment of sub-micrometer water droplets for disinfection of 3-d biological surfaces and production of liquid fertilizers. Water mist is generated by an ultrasonic nebulizer and is passed through a plasma generated by 29 kV positive pulses of ~5 ns duration in air. [G. Fridman, private communication, (2016)]

*Implants:* LTPs have long been used as a tool for improving surface interactions between materials and biological systems in various applications. Plasma processing in the health-care/biomaterials domain is currently experiencing spectacular growth [82].

Utilizing LTPs in biotechnology, medicine and agriculture is perhaps the most interdisciplinary and convergent of the major applications of LTPs. The overlap between LTPs and fundamental biology-related disciplines is broad and deep. These latter fields include biochemistry, microbiology, cell biology, medicine, botany, crop science, food science, agronomy and epidemiology. To truly leverage the potential of LTPs in areas of biotechnology, biomedicine and agriculture, there must be collaborative efforts between LTP researchers and researchers in these related areas. Some of these fields may be outside the traditional areas of NSF-funded engineering projects. This observation strengthens and emphasizes the imperative need for convergent research. Having said that, there are fundamental research issues that are also highly LTP-focused and that must be addressed to realize this potential.

IV.C.2. Synergies, Opportunities, Challenges

Applying LTPs to biotechnology has extraordinary promise in many application areas. Indeed, the field is being vigorously pursued throughout the world sponsored by many national level initiatives – except in the United States. Progress in this area in the US has been hindered by the difficulty of funding research that requires convergence of two or more sometimes radically different areas – *LTPs and medicine*, or *LTPs and agriculture*, or *LTPs and epidemiology*, or *LTPs and food science*. The response from agencies that fund LTPs is often – "We don't fund plant sciences." And the response from agencies that fund plant sciences is often – "We don't fund LTPs."

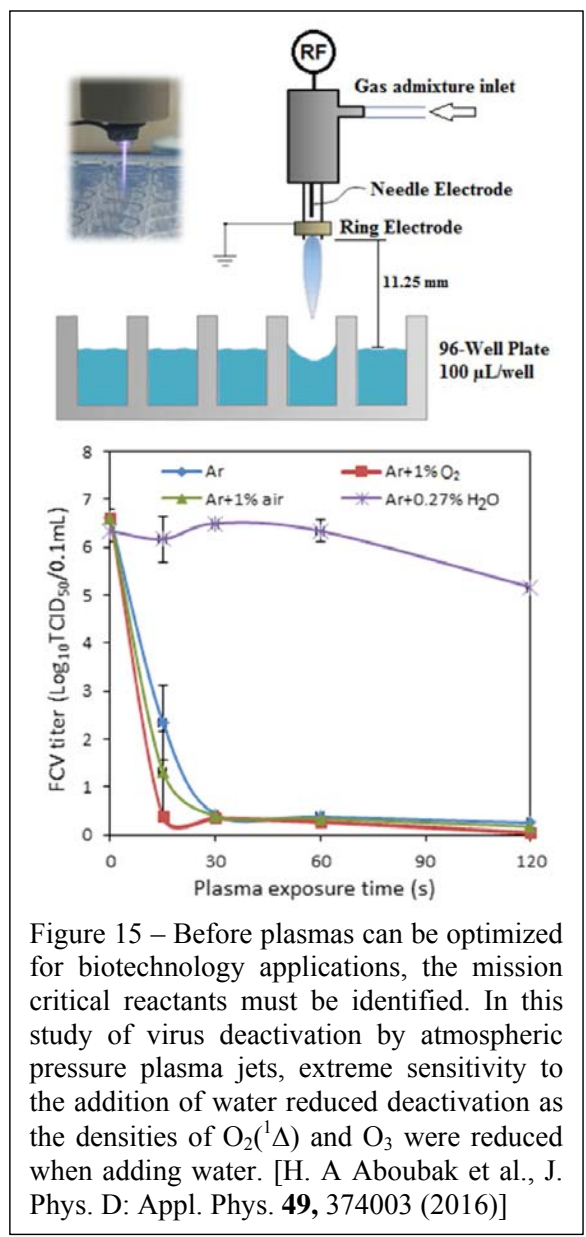

Figure 15 – Before plasmas can be optimized for biotechnology applications, the mission critical reactants must be identified. In this study of virus deactivation by atmospheric pressure plasma jets, extreme sensitivity to the addition of water reduced deactivation as the densities of $O_2(^1\Delta)$ and $O_3$ were reduced when adding water. [H. A Aboubak et al., J. Phys. D: Appl. Phys. **49,** 374003 (2016)]

To make progress in the science of LTPs in biotechnology and the food cycle, there must be close connectivity to the application. Collaboration is required to provide proper context for advances in the science of LTPs and to deliver rapid societal benefit through translational research. A typical scenario involves a demonstrated ability of LTPs to produce reactive fluxes which preferentially kill cancer cells while producing little or no harm to non-cancerous cells [83]. One LTP science issue is how to control the flow of power through the plasma to generate the precise reactive species that produce the therapeutic effect. The related question is how to adapt the ability to produce such reactive species to match the natural variability of patients and within a given strain of cancer. *How is success or failure detected?* The improved fundamental understanding enabling production of these reactive species translates to a plasma device that ideally kills can-cer cells while not killing healthy cells. Convergent research enables rapid feedback and collaboration between the LTP scientists and biologists.

Disease is a monumental global challenge, which is exacerbated by a growing resilience of various infectious disease threats. Antimicrobial resistance (AMR) is the inevitable development of resistance by bacteria and microbes to antimicrobial drugs [84]. It is estimated that more annual deaths result from AMR than from cancer. LTPs provide a possible route to combatting AMR. Plasmas, with their ability to controllably produce multimodal forms of antibacterial agents, can specifically target microbes in a manner that is less likely to develop resistance. In research to date, there is no evidence that microbes have developed any perceptible resistance to antimicrobial plasma treatment [85]. (See Fig. 16.)

Plasmas provide distinct advantages and complementary attributes compared with existing and emerging therapies, including reduced side effects. They offer focal therapies either solely or in conjunction with other modalities [86]. Through activating the immune system, they also offer potential to tackle disease and induce a protective mechanism [78]. In addition to human health benefits, there are direct connections to sustainability – reducing the over-use of antibiotics which lead to AMR, limiting the overuse of pharmaceuticals which accumulate in the environment and contaminate water supplies, cutting the use of antifungal agents in agriculture which accumulate in the environment and cause unintended harm to other species, and decreasing drugs in the food cycle (e.g., inoculating poultry and cattle) which eventually reach humans.

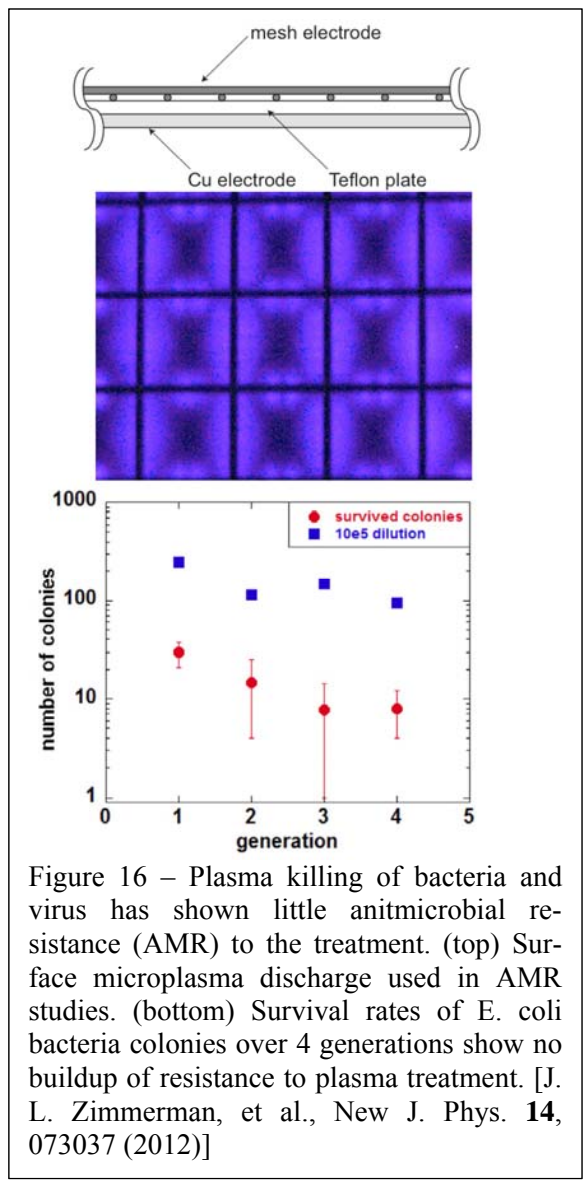

Figure 16 – Plasma killing of bacteria and virus has shown little anitmicrobial resistance (AMR) to the treatment. (top) Surface microplasma discharge used in AMR studies. (bottom) Survival rates of E. coli bacteria colonies over 4 generations show no buildup of resistance to plasma treatment. [J. L. Zimmerman, et al., New J. Phys. **14**, 073037 (2012)]

Atmospheric pressure LTPs, used in most plasma medical applications, have spatial and temporal inhomogeneities due to the inherent properties of the device [87]. Treated tissue surfaces then receive unpredictable variations in radicals, ions, electric fields and photons. These variations negatively impact the reproducibility of treatment and thus the ability to derive reaction pathways and mechanisms. There is clear interplay between flow dynamics, electrical parameters and chemical parameters of plasmas, which can be strategically employed to manipulate this non-equilibrium environment to produce desired reactive fluxes at the surface. For example, studies have shown that pulsed power waveforms can be used to tune electron velocity distributions and so modify radical fluxes incident onto surfaces. The research challenge is achieving this fine level of control in a reproducible manner at atmospheric pressure which can be tuned to meet specific patient needs.

This field, as well as all fields of biology, is challenged to equate cause and effect. Biological processes are exceedingly complex, and despite rapid progress in the field, it is beyond the state-of-the-art to *a priori* predict the metabolic reaction of a specific cell type to an arbitrary flux of reactive oxygen and nitrogen species (RONS). For the foreseeable future, there is need to develop the ability to accurately characterize the reactant fluxes (and reactant environments around cells) produced by the plasma to enable improved fundamental mechanistic understanding. Developing techniques to accurately measure and control the identities and concentrations of plasma produced species will enable us to determine energy partitioning and product branching ratios in key plasma chemical reactions.

Accomplishing these goals requires convergent research. Before optimizing the reactant fluxes, one needs to know which species' fluxes need optimizing [88]. This requires quantifying the effects and responses of biological systems resulting from well-characterized fluxes, including side effects at the intra- and extra-cellular level. The ability to observe and quantify the responses (effects) of plasma on biological systems is difficult because the diagnostic techniques must not impact the targeted parameter. Biological systems will generally respond to an outside stimulus generated by a diagnostic, and delicate biological processes can be easily perturbed. The measurements and data analysis will require a statistics-based approach to address the large number of required samples.

Developing a methodology or model system in which this cause-and-effect relationship can be established constitutes another major research challenge. This could be a representative but non-living system, such as liposomes filled with a scavenger in a pure water system. After characterizing the cause-and-effect in the transfer of reactivity from plasma-to-water-to-liposome, increasing levels of complexity could be added, eventually reaching the level of treating living cells.

Multiphase, multiscale models for plasmas, fluids and chemistry exist and are being further developed. The challenges in developing these models are discussed in Section IV.A. A unique challenge in developing integrated models of plasma-biological systems is quantifying the biological response with sets of partial differential equations. Despite impressive progress, physics-style models of biological systems are still in their infancy. In those models that do use differential equations (e.g., angiogenesis, simple wound healing and some parts of oncology) many parameters are often needed [89]. It is rare that these parameters can be independently estimated or measured. In many cases, the parameters are adjusted to obtain systematic agreement of model results with experiments. Such models are therefore not truly predictive. The dynamic range of relevant timescales is also incredibly large: collision frequencies within plasmas are on the order of nanoseconds; timescales of diffusion processes within cells can be on the order of hours; and timescales of genetic expression are at least multiple cell cycle periods. Accurate results in modeling may require microsecond time steps over many real hours, which discourages and negates a brute force approach. Progress requires an intelligent restriction of timescales to study each mechanism of interest.

Several classes of diagnostics are required to resolve transport from the plasma to and into the biological system. Diagnostics to fully characterize even the relatively simple plasma-to-water-to-liposome system are not currently available. The plasma source and fluxes to the liquid must be characterized, as must the plasma produced reactivities in the liquid and the liposome. These diagnostics should be performed in as close to real-time as possible to capture important short-lived species and their effects. These diagnostics in large part do not exist today. Modeling challenges, even in the simple plasma-to-water-to-liposome system, are also formidable. The plasma produced reactivity passes through at least three phase boundaries (plasma to water, water to liposome membrane, then liposome membrane to interior). The precise transport mechanisms through these phase boundaries are poorly understood. Although significant progress is being made [90][91], no model is currently capable of addressing all the critical system parameters in an integrated manner.

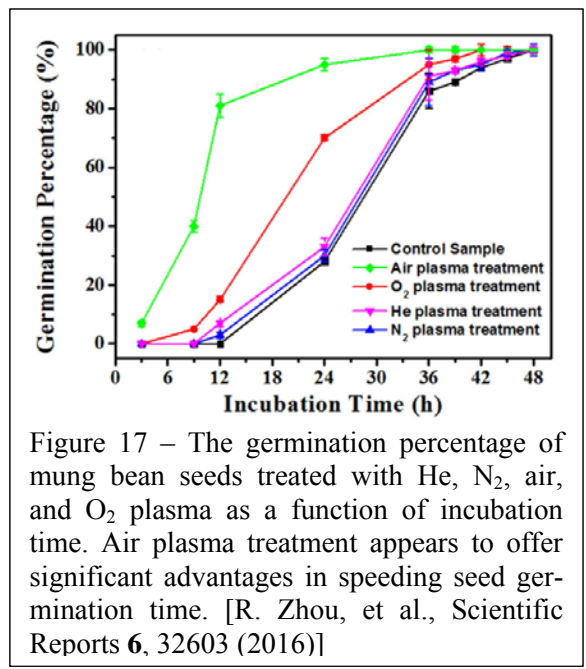

Figure 17 – The germination percentage of mung bean seeds treated with He, $N_2$, air, and $O_2$ plasma as a function of incubation time. Air plasma treatment appears to offer significant advantages in speeding seed germination time. [R. Zhou, et al., Scientific Reports **6**, 32603 (2016)]

LTPs are now being intensively investigated for beneficial uses in agriculture and food applications [71][92]. For example, LTP treatment of seeds has been shown to increase crop yields in multiple studies, both by increasing the fraction of germinating seeds, decreasing germination time and by enhancing subsequent crop growth [93]. For example, a dramatic shortening of mung bean seed germination time due to air plasma treatment has been demonstrated [94]. (See Fig. 17.)

One promising application of LTP to seed treatment involves eliminating pathogenic fungi from seed surfaces [95]. Using LTP in this manner has direct benefit to the environment by reducing or eliminating the use of environmentally damaging liquid treatments. For example, Dobrin et al. state that LTP has a "…positive effect on wheat early growth. Due to its advantages (uniform treatment, no destruction of seeds, no requirement for chemicals), plasma might become an effective alternative to traditional pre-sowing seed treatment used in agriculture" [96]. One possible mechanism by which plasma treatment positively affects seeds is increasing water absorption. In an intriguing preliminary study of epigenetic effects of plasma on treated seeds, growth was promoted in subsequent generations of untreated daughter plants [97].

LTP treatment of plants offers the advantage that plants are easier and less expensive to study than animals or even mammalian cell cultures. Since many (but not all) plasma-biology interactions are expected to be similar for plants as for other forms of aerobic living matter, studies of LTP-plant interactions could help accelerate research in the entire area of plasma biology.

Recent research demonstrates the potential of air plasma to generate nitrogen-based fertilizers in relatively small-scale mini-plants, thus replacing or reducing the demand for nitrogen fertilizers from large-scale, centralized (Haber-Bosch) $NH_3$ manufacturing plants. Suggestions have been made to combine air plasma with heterogeneous catalysis to improve air plasma energy efficiency [98]. A variation on this theme is to use air LTP-generated $NO_x$, dissolved in water, to acidify organic waste such as animal manure. Bacterial degradation of organic waste generates copious quantities of $NH_3$, and acidification via aqueous $HNO_3$ captures this otherwise fugitive (and environmentally damaging) reactive nitrogen as $NH_4NO_3$, thus increasing the N-content of organic fertilizer [99]. Air LTP mini-plants would most likely be located near farms and powered using local renewable energy resources such as wind and solar.

Soil treatment and sterilization is another promising agricultural application of LTPs [100].

For example, plasma-generated ozone can serve as a pesticide against nematodes (worms) while potentially preserving beneficial bacteria and at the same time promote plant growth (radishes in their study), thereby acting as a kind of growth-promoting fertilizer [101]. Note that currently, nematodes are eliminated from many agricultural soils using methyl bromide ($CH_3Br$) fumigation, a process that must be replaced for environmental reasons [102].

LTP-based food disinfection has received a great deal of attention lately [103]. LTPs have successfully been used to sterilize the surfaces of many varieties of food and food products, including egg shells [104], fruits and seeds [105][106], and meat and cheese [107]. (See Fig. 18.) Gas plasmas created above a liquid (water, milk and fruit juices) have been used to inactivate pathogens at moderate temperatures and short treatment times [108].

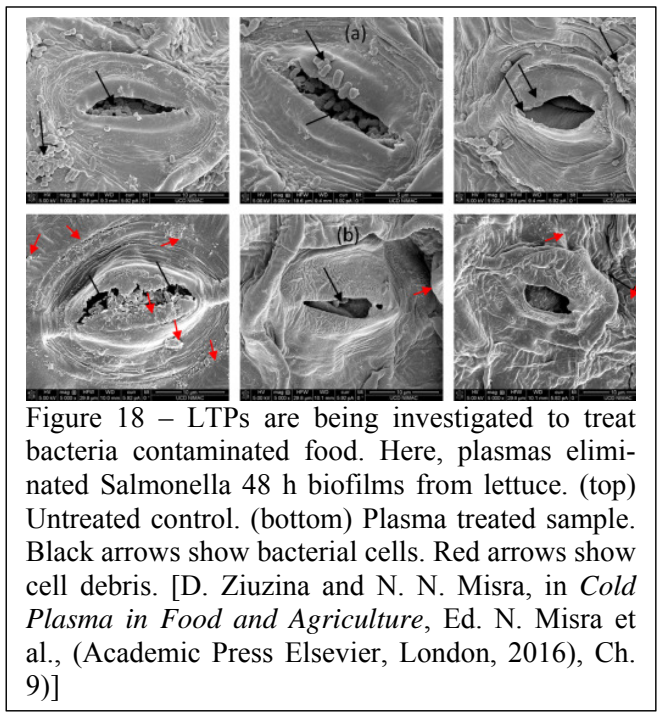

Figure 18 – LTPs are being investigated to treat bacteria contaminated food. Here, plasmas eliminated Salmonella 48 h biofilms from lettuce. (top) Untreated control. (bottom) Plasma treated sample. Black arrows show bacterial cells. Red arrows show cell debris. [D. Ziuzina and N. N. Misra, in *Cold Plasma in Food and Agriculture*, Ed. N. Misra et al., (Academic Press Elsevier, London, 2016), Ch. 9)]

Potential advantages of LTP food disinfection include: (i) relatively high microbial inactivation efficiency at near-room temperatures, (ii) *in situ*, on-demand, compatibility with food packaging, (iii) rapidly acting active agents having few, if any, negative effects on most food products, (iv) food preservative use, (v) generally benign environmental effects with few or no food residues, and (vi) high energy efficiency [109]. Research challenges persist in areas of control addressing variability, matching plasma source and production of radicals with intended applications, and understanding the underlying plasma chemical processes that penetrate the porous skins of fruits and vegetables or the surface layers of meat, to react with bacteria.

To achieve breakthroughs in LTPs applied to food and agriculture, multiple challenges must be overcome. These include the need for close collaboration between LTP experts and plant pathologists, agronomists, and other agricultural and food specialists. In addition, LTP applications in this area are very much in an early stage, and it is unclear what plasma species and types of sources will work best. Furthermore, appropriate plasma diagnostic, modeling and control technologies must be explored and novel LTP technologies invented to realize the many possibilities, especially at the scales needed for worldwide impact on sustainability.

Every instance of plasma based processes replacing chemical based processes in biotechnology and the food cycle will reduce the environmental impacts associated with generating chemicals and waste disposal. If the plasmas are produced with renewable electricity, the environmental impact is even smaller. The resulting improvements in sustainability and environmental stewardships will help enable a FBRE.

IV.C.3.    Summary of Research Challenges

As demonstrated above, the research issues dominating LTP science in biotechnology and the food cycle are broadly encompassing and invariably overlap with biology. Two wide-ranging plasma centric categories of research challenges include the following.

- *Controlling energetic particle distributions* – Identifying how the non-equilibrium LTP environment expressed through control of the electron energy distributions can be manipulated to deliver reactive species that trigger specific bio-chemical processes. Potential applications include:

  - DNA damage by low energy electrons;
  - Biological reactions involving UV light;
  - Wound healing through production and control of reactive nitrogen species (RNS);
  - Cancer treatment through production and control of reactive oxygen species (ROS);
  - Skin treatment, and
  - Antibacterial protection.

- *Developing flexible, individualized plasma systems to address biological variability* – Developing personalized plasma sources and treatments that possess sufficient flexibility and control to account for the intrinsic variability of biological systems: organism-to-organism, patient-to-patient, plant-to-plant. Elements of this research include:

  - Assessing critical plasma parameters that influence consistency and overcome biological variance (close to patient models) using real-time feedback control;
  - Manipulating spatial and temporal inhomogeneities in LTPs, and
  - Determining the effects of treatment on organoleptic properties and chemical components in food.

### IV.D.     *Synthesis and Modification of Materials*

IV.D.1.     Introduction

Solving the grand challenges facing society, including protecting human health and addressing the *food, energy, water nexus*, requires developing advanced materials. For example, advanced materials are needed to modify surfaces for biocompatibility of implants, replace glass in photo-voltaic modules with coated polymers, create new membranes for water purification and provide sustainable materials for energy storage. To achieve the next breakthrough, these materials require precise control of properties such as composition, interfaces, structure, morphology and surface termination, while being composed from sustainable, earth-abundant elements and compatible with large-scale manufacturing. Plasmas offer the ability to combine materials that are not compatible at high temperatures. For example, depositing aluminum layers on silicon is impossible without plasma.

Our ability to dynamically tailor electron and ion energy distributions, fluxes of radicals, gas-phase reactive chemistry and electric fields within the LTP environment provides an unprecedented opportunity to address these grand challenges. LTPs' unique non-equilibrium properties have already been successfully utilized to meet the needs of industries in areas such as semiconductor processing, lighting and advanced coatings. LTPs are an excellent candidate to meet the materials requirements for a sustainable future. Nevertheless, a large knowledge gap exists between the materials and properties we desire and designing a plasma process to realize them.

IV.D.2.     Synergies, Opportunities, Challenges

Plasma based materials synthesis is critical to the energy and sustainability sectors. Particularly relevant is the utilization of plasmas for both the synthesis and activation (via surface modification) of catalyst nanoparticles used in chemical conversion. Plasma-based materials synthesis also enables wiser use of resources. Replacement of materials that are expensive, toxic or thermally unstable (for instance, displacing CdS quantum dots or precious metal plasmonics) may be enabled by plasma synthesis. Energy storage applications also require new materials, many of which will rely on plasma synthesis. For example, advanced batteries require unique nanostructured materials capable of being economically deposited on non-planar substrates. The unique properties of microplasmas are already making inroads to performing such depositions [110]. (See Fig. 19.)

Plasma materials processing can generally be divided into two regimes – low pressure (typically less than a few Torr) and high pressure (typically 1 atmosphere). This division derives from practical technological considerations: it is difficult to engineer systems that operate at intermediate pressures and can economically scale to mass production.

*Low Pressure Materials Processing*

Low pressure plasma materials processing has achieved unparalleled success in developing processes for microelectronic fabrication and now, nanoelectronics fabrication. LTP-based transfer of photoresist patterns into other materials (e.g., Si, $SiO_2$, $Si_3N_4$, $HfO_2$, GaAs) has been remarkably successful in enabling production of nanometer scale devices [111]. Without this capability, most electronics devices and applications that have emerged over the last 40 years, including

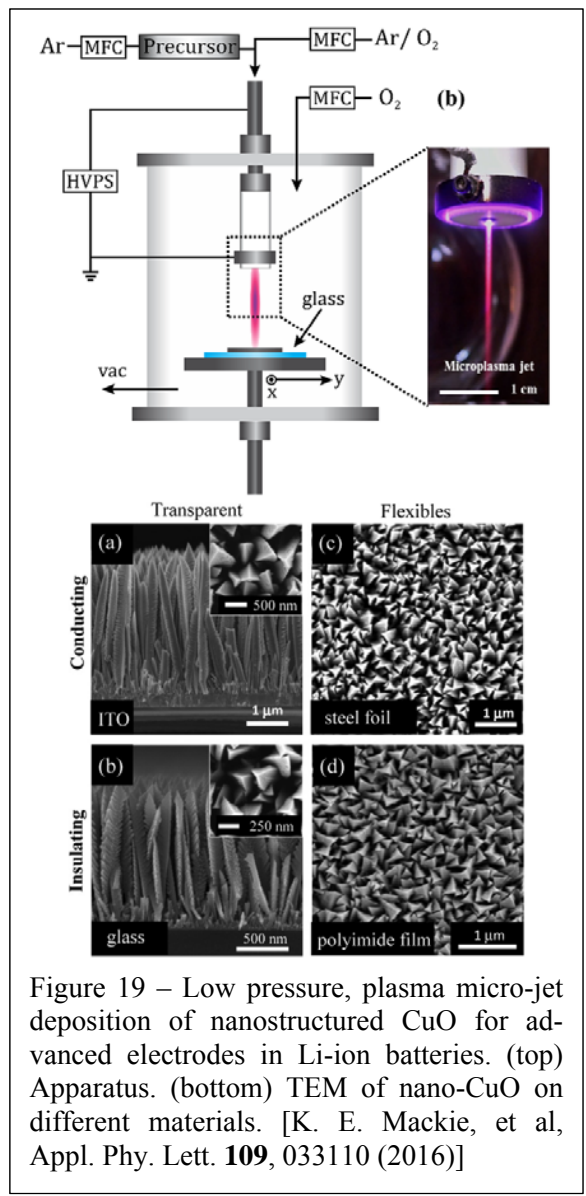

Figure 19 – Low pressure, plasma micro-jet deposition of nanostructured CuO for advanced electrodes in Li-ion batteries. (top) Apparatus. (bottom) TEM of nano-CuO on different materials. [K. E. Mackie, et al, Appl. Phy. Lett. **109**, 033110 (2016)]

personal computers, smart phones, digital cameras, to name just a few – essentially most of the current electronics and indeed Moore's Law of the semiconductor industry – would not have been possible. The continuing drive to decrease feature sizes in nanomanufacturing has led to plasma etching techniques that can provide control at atomistic length scales [5]. The plasma based solution is atomic layer etching (ALE) techniques [112][113]. After nearly 2-decades of visioning, research and engineering, plasma based ALE is now being implemented in manufacturing.

LTP-based synthesis of advanced materials has played a similar transforming role in multiple industries, including clean energy. Absent the breakthrough demonstration of plasma enhanced chemical vapor deposition (PECVD) doping of amorphous silicon with boron and phosphorus in 1975 by Spear and Lecomber [114], amorphous solar cell technology, liquid crystal displays and other ubiquitous products would not have been economically possible. This breakthrough demonstrated how the non-equilibrium nature of LTPs was essential in developing a new materi-al that has had untold positive impacts on the environment by enabling inexpensive, clean solar electrical power generation.

Current thin film coating applications require layers that are either only several atoms thick [115], consist of nanoscale materials (e.g., nanocrystalline materials [116][117]), are extremely robust in aggressive environments, such as coatings of extremely tough and hard steels for cutting tools in machining operations [118], or have other functionalities [119]. Plasma-assisted deposition based on low ion energy stimulation of the surface has been used for plasma-assisted atomic layer deposition (ALD), the analog to ALE [115]. Dusty plasmas have been utilized for the synthesis and modification of nanoparticles including metals, semiconductors and ceramics, for which size, structure, chemical composition, and other aspects can be controlled [120]. High power impulse magnetron sputtering (HiPIMS) is a recent development of the physical sputtering technique that can produce high quality materials at high rate [121]. HiPIMS is based on application of high power in short pulses that produce a dense plasma that is transient in nature that results in improved control of film quality, such as film density and adhesion to the substrate. However, the transient nature of the plasma and the complex transport dynamics of charged particles in a magnetic field used for HiPIMS creates a plasma environment that is strongly time and position dependent [122].

State-of-the-art plasma processes for materials synthesis requires controlling the energy and composition of the species fluxes from the plasma to the material, including ions, neutrals, photons and electrons. Steady-state plasma systems provided this level of control for early material synthesis where structures were large and composition was not critical. As structures became smaller, selectivity became more critical and composition control became paramount, advanced plasma processes and reactors emerged. These new processes often use pulsed plasma power and/or biasing of the substrate, pulses of reactive gases or spatial separation of plasma generation and material application. These techniques achieve improved control of surface reactions and of the overall outcome of plasma-induced synthesis, etching or surface modification reactions.

It is expected that greater degrees of control will be required for materials used in environmental, energy and healthcare applications. Although the challenges of plasma control for nanoelectronics fabrication will persist far into the future, these materials, once formed, are not chemically reactive. In contrast, the vast majority of materials for environmental, energy and healthcare applications are required to be chemically reactive, from selectively reacting catalysts, to nanostructures which enable binding to select chemical species for detectors to cell-adhering (or not-adhering) surfaces. The ability to customize both the physical structure and the chemical reactivity of these materials, places an entirely higher level of expectation on the ability of LTPs to deliver the required fluxes to produce this functionality.

There are also important environmental uses for plasma produced functionality on materials spanning very large areas. For example, the buildup of biofilms on ship hulls can increase drag by 20%, and the increase in drag by barnacles can exceed 60% [123]. Drag increases fuel consumption, which translates to more emission of harmful gases into the atmosphere. Due to negative environmental consequences, chemicals are not a desirable solution. A previous chemically based solution to prevent fouling of ship hulls precipitated a biological crisis when the chemical accumulated in marine life [124]. Modern antifouling solutions rely on chemically functionalizing the surface with benign hydrophilic polymer brushes or creating nanoscale roughness on the surface, both of which have been performed on small surfaces (tens of cm) by LTPs [125]. The plasma science challenge is scaling these laboratory processes to application on surfaces the size of oil-tanker hulls.

Historically, nanoparticles in plasmas have gone from being unwanted contamination to a method of fabricating advanced materials. Plasmas used in the semiconductor industry for the deposition and etching of thin films produced nanoparticles by a process known as homogeneous nucleation. Referred to as dust, these nanoparticles could compromise the thin film devices being fabricated. Early research on dusty plasmas focused on preventing particle formation and contamination of wafers. With the advent of nanotechnology in the 1990s, research on nanoparticle formation in plasmas shifted from mitigation to purposeful enhancement and scale-up of nanoparticle production. Today, plasma synthesis of nanoparticles is a relatively well-established area of scientific research, with basic reactor strategies that are distinct from other processes such as thin film etching and deposition, diagnostics aimed at monitoring and understanding particle nucleation and growth, and applications of plasma-produced nanoparticles in electronics, catalysis and energy [32]. Although this discussion is under the low pressure plasma heading, nanoparticle synthesis is also being investigated at atmospheric pressures using microplasmas.

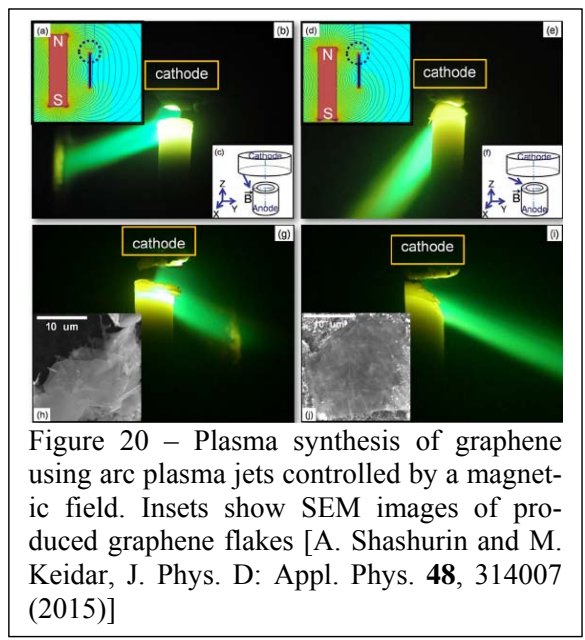

Figure 20 – Plasma synthesis of graphene using arc plasma jets controlled by a magnetic field. Insets show SEM images of produced graphene flakes [A. Shashurin and M. Keidar, J. Phys. D: Appl. Phys. **48**, 314007 (2015)]

The landscape of nanomaterials synthesized by plasmas is at the early stages. (See Fig. 20.) The majority of work on this topic has focused on silicon which has a long history in the plasma community. Only recently has research extended to other materials including metals [126], alloys [127], carbon structures such as graphene [128], and doped semiconductors [129].These materials introduce new questions that couple the fundamental processes in the plasma and the properties of the materials, which must be controlled or tuned for specific applications, such as high efficiency solar cells.

This change in emphasis in nanoparticle-containing plasmas from contamination-free-manufacturing to nanoparticle synthesis created additional scientific questions and technological opportunities. Unlike conventional thin film etching and deposition, where the plasma is homogenous and surfaces are on the boundaries, nanoparticle-forming plasmas are highly complex multiphase systems composed of the gaseous plasma components mixed with atomic- to nanoscale clusters, nanoscale particles, and nano- to micro-scale particles and aggregates. In addition to size distribution, there is a wide spectrum of particle compositions, structures and charge states. The plasma-nanoparticle interaction takes on a higher level of complexity. For example, local changes in plasma properties feed back to the growth of the nanoparticles that in turn affect the plasma through changes in their charge state, secondary emission properties and chemical catalytic properties. To model such systems, small clusters must be treated differently than larger nanoparticles, as the charge state, structure and other properties (e.g., electronic, optical, magnetic, etc.) can be vastly different. Diagnostics must characterize both the plasma phase and the nanoparticles, encompassing both plasma and aerosol science. The consequences of adding chemical reactions to dusty-plasma-physics also remains relatively unexplored.

*High Pressure Materials Processing*

Although traditionally not classified as non-equilibrium plasmas, thermal plasmas are unique sources of thermal energy and chemically active species. Thermal plasma processes have a proven track record for unique versatility in a wide variety of fields, including metal cutting and welding, extractive metallurgy, physical and chemical vapor deposition, particle and chemical synthesis, biomass gasification and waste treatment. (See Fig. 21.) A distinct high-pressure plasma technology, thermal plasma spray is one of the most versatile techniques for applying protective and functional coatings, such as thermal barriers and wear- and corrosion-resistant coatings using a wide variety of materials, especially refractory materials (e.g., oxides, molybdenum), in industries as diverse as transportation, energy, materials extraction and processing, biomedical and electronics [130].

At the heart of thermal plasma processing is the plasma source [131]. The continuous development of plasma sources, such as plasma torches, has been possible due to improved understanding of high-pressure plasma generation, confinement, and interaction with processing media. The past record of success in this regard is impressive. Fundamental investigations in the mid-1960s of the dynamics of a confined arc interacting with cold gas flow [132] led to standard and advanced torch designs [133][134][135][136]. These advances were facilitated by continuous advancement of electrical, optical, thermal and acoustic plasma source diagnostics combined with high-fidelity computational simulations, which now use nonequilibrium plasma flow models [137].

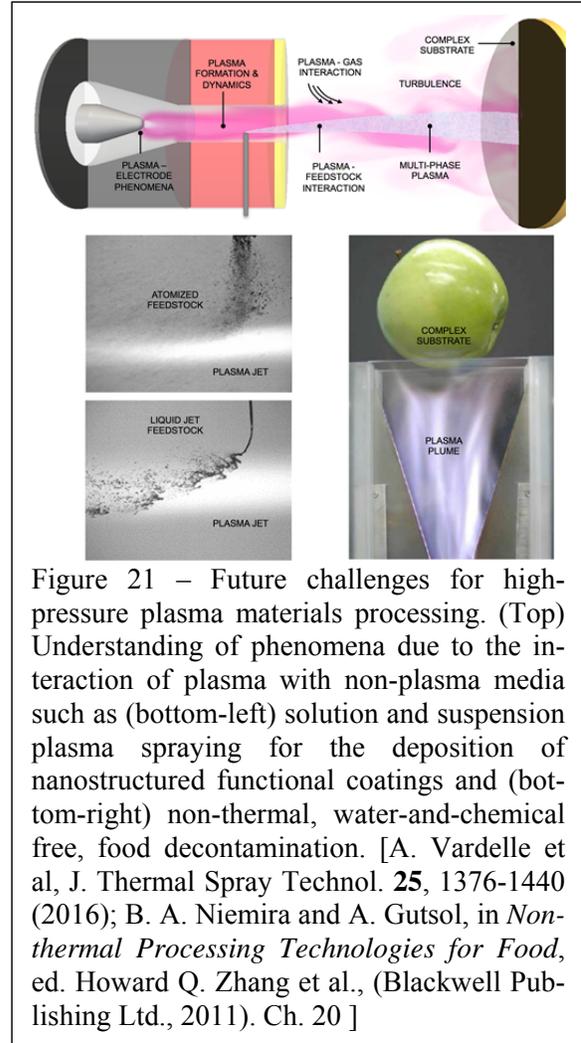

Figure 21 – Future challenges for high-pressure plasma materials processing. (Top) Understanding of phenomena due to the interaction of plasma with non-plasma media such as (bottom-left) solution and suspension plasma spraying for the deposition of nanostructured functional coatings and (bottom-right) non-thermal, water-and-chemical free, food decontamination. [A. Vardelle et al, J. Thermal Spray Technol. **25**, 1376-1440 (2016); B. A. Niemira and A. Gutsol, in *Non-thermal Processing Technologies for Food*, ed. Howard Q. Zhang et al., (Blackwell Publishing Ltd., 2011). Ch. 20 ]

The research challenges for thermal plasma material modification are driven by emerging applications. These challenges include the interaction of plasma with novel media, from complex substrates (e.g., nanostructured surfaces) to new types of feedstock (e.g., liquid streams, droplets, bio-feedstocks, municipal solid waste MSW). In each case, tuning of plasma characteristics and behavior is required to account for the unique response of the feedstocks, from plasma stability to plasma reactivity and energy density. For example, using liquid feedstock imposes new challenges to ensure uniform processing and requires understanding the interaction of plasma and liquid streams, the transfer of species and energy across a liquid interphase and the effect of high vapor pressures within the plasma, among other phenomena.

High-pressure plasma research has traditionally been performed in a weakly-coupled manner; the study of fundamental plasma phenomena is often separated from phenomena associated with the processing agent (e.g., injected powder or droplets). Emerging and future technological ap-plications will require strongly-coupled research strategies in which the plasma and the processing agent are studied concurrently while combining experimental and computational methods in a coordinated manner. This concern extends to environmental applications (such as biomass and MSW conversion) where the feedstock will have tremendous variability.

Experimental challenges include multi-dimensional diagnostics with high temporal- and spatial-resolution capable of capturing sub-micron features such as initiation of instabilities or the establishment of coherent (long-lived) flow structures [138], non-intrusive methods for probing the degree of non-equilibrium of the plasma to determine electron temperature and reactive species concentrations, and methods applicable to multiphase systems (plasma-liquid, plasma-solid). Computational challenges include robust methods suitable for addressing shear instabilities leading to turbulence and arcing instabilities, combined with non-equilibrium plasma models that can accommodate comprehensive kinetic mechanisms.

An entirely new field of processing non-equilibrium materials using atmospheric pressure plasmas is emerging for deposition of functionalized coatings [139]. These systems will soon extend to fabricating coatings having complex nanoscale structures on flexible materials, such as polymers that are compatible with roll-to-roll web-type processing. Successful development of such processes will enable, for example, extremely inexpensive solar-cell production on flexible substrates in the same manner that plastic wrap is functionalized today.

IV.D.3.     Summary of Research Challenges

Plasma-surface interactions dominate the production of new materials and so are critical in producing new functionality. These plasma-surface interactions present a unifying challenge across the focus areas, and are critically important given that product here is usually a solid material. Thus, understanding, controlling and predicting plasma-surface interactions is the most critical scientific challenge that will allow us to use LTPs to address the materials needs of sustainability. To address this challenge, convergent research must focus on plasma and materials diagnostics and modeling to understand the fundamental phenomena, tightly enmeshed with the application areas critical to furthering sustainability. These are detailed below.

- *Developing the diagnostics, procedures and models required to understand and control plasma-surface interactions*. The following diagnostics and methods will enable scientists to characterize and control these interactions:
  - Detailed/real-time/*in-situ* diagnostics of both the plasma phase and the evolving surface;
  - Computational tools to bridge the gap between gas phase reactive species and dynamic surface chemistry and structure, and
  - Methods (waveforms, frequencies, reactor geometries, gas composition, bias) to control plasma properties such as electron and ion distribution functions, radical and photon fluxes, and potential bias.

- *Tackling application areas critical to a sustainable future*. The following application areas present immediate challenges to advancing sustainability:
  - Large-area two-dimensional material processing at low thermal budget on arbitrary substrates that are potentially sensitive to high temperatures for photovoltaics, permeable membranes for water purification, and antifouling coatings on ships and the inside of pipes;
  - Thermal and chemical barrier coatings for high efficiency engines, turbines and chemical

processing;
- Energy conversion and storage, and chemical conversion materials – including nanostructured materials for batteries, customized catalysts for chemical conversation, and
- Optically active nanoparticles (quantum dots, plasmonics) with applications in photonics (PV, sensors), bio and health care (tagging and theranostics).

## V. Concluding Remarks – Looking Ahead

The field of low temperature plasmas (LTPs) promises exciting solutions to some of the world's most pressing concerns, particularly with regards to sustainability and a *future based on renewable electricity (FBRE)*. While the challenges presented here are daunting, they are not out of reach. As mentioned throughout this report, convergent research is a fundamental condition for meeting these challenges – and this is a hallmark of LTPs. In the context of LTPs, convergence includes a wide variety of scientific disciplines plus a broad range of engineering specialties performing collaborative application-motivated research focused on achieving the potential of plasma enabled sustainability.

While this convergent research in LTPs is currently occurring through the individual efforts of members of the community, accomplishing the challenges discussed in this report requires more deliberate action. Achieving the stated goals will only be possible with a dedicated programmatic home for LTP research. Moving forward, establishing an LTP program that is empowered to pursue convergent research at the National Science Foundation would achieve many of these goals. Given the translational nature of this research, the Directorate of Engineering would be an appropriate host. Establishing a programmatic home for LTP research will enable the US to remain at the forefront of efforts to achieve sustainability, and to be at the forefront of LTP science and engineering.

List of Figures

# Appendix A: Workshop Agenda

## NSF Low Temperature Plasma Workshop: Agenda
NSF Headquarters, Stafford Place 2 Building, Rooms 555 & 575

| Day 1 – August 22 | |
|---|---|
| 7:45 – 8:15 am | Registration |
| 8:15 – 8:30 am | **Introduction to Workshop and Goals**: Mark J. Kushner |
| 8:30 – 8:35 am | **Welcome From NSF**: Dr. Grace Wang, Asst. Director for Engineering |
| 8:35 - 9:15 am | **Plenary Overview of Research Challenges** (Chair: Mark J. Kushner) <br> Richard van de Sanden (Dutch Institute for Fundamental Energy Research, Netherlands) |
| | **Topical Overviews and Prospective Challenges** (Chair: Selma Mededovic Thagard) |
| 9:15 – 9:45 am | *Multiphase Systems – Liquids and Aerosols*: <br> Peter Bruggeman (U Minnesota) |
| 9:45– 10:15 am | *Energy and the Environment:* <br> Hyun-Ha Kim (National Institute of Advanced Industrial Science and Technology, Japan) |
| 10:15– 10:30 am | Break |
| 10:30–11:00 am | *Synthesis and Modification of Materials*: <br> Gottlieb Oehrlein (U Maryland) |
| 11:00– 11:30 am | *Biotechnology and Food Cycle*: <br> David B. Graves (U California-Berkeley) |
| 11:30 –12:00 pm | **Mini-presentations (5 mins) by Workshop Attendees** (Chair: R. Mohan Sankaran) |
| | Elijah Thimsen: *Increasing the productivity of low temperature plasma reactors* | Alexey Shashurin: *Microwave scattering for diagnostics of atm pressure microplasmas* |
| | Azer Yalin: *New approaches to tailor and control laser induced air plasmas* | Scott Baalrud: *Plasma-Boundary Interactions in Support of 21$^{st}$ Century Engineering* |
| | Michael Keidar: *Understanding mechanism of plasma interaction with cancer & normal cells* | |
| 12:00 – 1:30 pm | Lunch - |
| 1:30 – 1:35 | **NSF Perspectives:** Dr. Triantafillos (Lakis) Mountziaris |
| 1:35 – 1:40 | **Introduction to Breakout Sessions:** (Mark Kushner) |
| 1:40 – 2:30 pm | **Breakout Session 1A** |
| | **Multiphase Systems:** Discussion Leaders – David Staack and Steven Shannon |
| | **Energy & Environment**: Discussion Leaders – Juan Pablo Telles and Uwe Czarnetzki |
| 2:35 – 3:25 pm | **Breakout Session 1B:** Multiphase Systems / Energy & Environment) |
| 3:25 – 3:40 pm | Break |
| 3:40 – 4:30 pm | **Breakout Session 2A** |

|  |  |  |
|---|---|---|
|  | **Materials:** Discussion Leaders – Rebecca Anthony and Lorenzo Mangolini | |
|  | **Biotechnology:** Discussion Leaders – Sylwia Ptasinska and Deborah O'Connell | |
| 4:35 – 5:25 pm | **Breakout Session 2B**: Materials / Biotechnology | |
| 5:30 – 5:45 pm | **Summary and Evening Assignments:** Mohan Sankaran | |
| **Day 2 – August 23** | | |
| 8:00 – 8:10 am | **Introductory Remarks** : Mohan Sankaran | |
| 8:10 – 8:40 am | **Mini-presentations (5 mins) by Workshop Attendees** (Chair: R. Mohan Sankaran) | |
|  | Igor Adamovich: *Characterization and Modeling of* LTPs *above and at a Phase Boundary* | Igor Kaganovich: *The Physics and Applications of Coherent Plasma Structures* |
|  | Sung-O Kim: *Optical fiber based microplasma devices* | Ed Barnat: *Diagnostics to facilitate research and enable discovery in* LTP *science* |
|  | Sergey Macheret: *Nanoscale spatiotemporal control of plasmas for energy, biotechnology, and materials synthesis* |  |
|  | **Preliminary Report and Discussion** (Chair: David B. Graves) | |
| 8:40 – 9:10 am | **Multiphase Systems**: David Staack and Steven Shannon | |
| 9:15 – 9:45 am | **Energy & Environment**: Juan Pablo Telles and Uwe Czarnetzki | |
| 9:50 – 10:20 am | **Synthesis and Modification of Materials:** Rebecca Anthony and Lorenzo Mangolini | |
| 10:25 –10:55 am | **Biotechnology and Food Cycle**: Sylwia Ptasinska and Deborah O'Connell | |
| 10:55 –11:15 am | Break | |
| 11:15 – noon | **Discussion and Writing Assignments** (Mededovic-Thagard, Sankaran, Kushner) | |
| noon – 1:00 pm | Lunch  -Working- | |
| 1:00 – 2:20 pm | Writing Sessions and discussions (All Topics) | |
| 2:20 – 2:35 pm | Break | |
|  | **Summaries**: (Chair: Selma Mededovic Thagard) | |
| 2:35 - 2:45 pm | **Multiphase Systems:** David Staack and Steven Shannon | |
| 2:50 - 3:00 pm | **Energy & Environment:** Juan Pablo Telles and Uwe Czarnetzki | |
| 3:05 – 3:10 pm | **Materials:** Rebecca Anthony and Lorenzo Mangolini | |
| 3:15 – 3:25 pm | **Biotechnology:** Sylwia Ptasinska and Deborah O'Connell | |
| 3:30 – 3:40 pm | **Concluding Remarks and Path Forward for Report:** (Mark Kushner) | |
| 3:40 pm | Adjourn | |

## Appendix B: Workshop Attendees

| Name | Institution |
|---|---|
| Igor Adamovich | Ohio State University |
| Sumit Agarwal | Colorado School of Mines |
| Rebecca Anthony | Michigan State University |
| Scott Baalrud | University of Iowa |
| Michael P. Bakas | Army Research Office |
| Sally Bane | Purdue University |
| Edward Barnat | Sandia National Laboratory |
| Curt Bolton | Department of Energy Office of Fusion Energy Science |
| Peter J. Bruggeman | University of Minnesota |
| Uwe Czarnetzki | Ruhr University - Bochum, Germany |
| Vincent Donnelly | University of Houston |
| J. Gary Eden | University of Illinois at Urbana/Champaign |
| John E. Foster | University of Michigan |
| Gregory Fridman | Drexel University |
| Steven L. Girshick | University of Minnesota |
| Matthew Goeckner | University of Texas at Dallas |
| Michael Gordon | University of California at Santa Barbara |
| David B. Graves | University of California at Berkeley |
| Chunqi Jiang | Old Dominion University |
| Igor Kaganovich | Princeton Plasma Physics Laboratory |
| Michael Keidar | George Washington University |
| Hyun-Ha Kim | National Inst. Advanced Industrial Science & Technology, Japan |
| Sung-O Kim | New York Institute of Technology |
| Vladimir Kolobov | CFD Research Corp. |
| Michael Kong | Old Dominion University |
| Uwe Kortshagen | University of Minnesota |
| Mark J. Kushner | University of Michigan |
| Vyacheslav Lukin | National Science Foundation |
| Sergey Macheret | Purdue University |
| Lorenzo Mangolini | University of California at Riverside |
| Selma Mededovic-Thagard | Clarkson University |
| Triantafillos Mountziaris | National Science Foundation |
| Deborah O'Connell | York University, United Kingdom |
| Gottlieb Oehrlein | University of Maryland |
| Nirmol Podder | Department of Energy Office of Fusion Energy Science |
| Sylwia Ptasinska | University of Notre Dame |
| Mohan Sankaran | Case Western Reserve University |
| Steven C. Shannon | North Carolina State University |
| Alexey Shashurin | Purdue University |
| David Staack | Texas A&M University |
| Andrey Y. Starikovskiy | Princeton University |
| David Stepp | Army Research Office |
| Elijah Thimsen | Washington University |
| Juan Pablo Trelles | University of Massachusetts at Lowell |
| Richard van de Sanden | Dutch Institute for Fundamental Energy Research, Netherlands |
| John P. Verboncoeur | Michigan State University |
| Mitchell Walker | Georgia Institute of Technology |
| Scott Walton | Naval Research Laboratory |
| Azer Yalin | Colorado State University |